\documentclass[letterpaper,11pt]{article}
\usepackage{color}
\usepackage{grffile}
\usepackage{ctable}
\usepackage{amssymb}
\usepackage{amsmath}
\usepackage[dvips]{epsfig}
\usepackage[small]{caption}
\usepackage{subfig}
\usepackage{graphicx}
\usepackage[letterpaper,text={16cm,23cm},centering]{geometry}

\newtheorem{Lemma}{Lemma}[section]
\newtheorem{Theorem}{Theorem}
\newtheorem{Proposition}[Lemma]{Proposition}
\newtheorem{Corollary}[Lemma]{Corollary}
\newtheorem{Remark}[Lemma]{Remark}
\newtheorem{Definition}[Lemma]{Definition}

\newtheorem{Example}[Lemma]{Example}

\newenvironment{Proof}%
 {\begin{trivlist} \item[]{\bf Proof. }}%
 {\hspace*{\fill}$\rule{.4\baselineskip}{.4\baselineskip}$\end{trivlist}}

\newenvironment{Acknowledgment}%
 {\begin{trivlist}\item[]\textbf{Acknowledgments.}}{\end{trivlist}}

 {\begin{trivlist} \item[]{\bf\color{red} Comment. }}%
 {\color{red}\hspace*{\fill}$\rule{.4\baselineskip}{.4\baselineskip}$\end{trivlist}}

\makeatletter\@addtoreset{figure}{section}\makeatother

\makeatletter \@addtoreset{equation}{section} \makeatother

\newcommand{\R}{\mathbb{R}}
\newcommand{\C}{\mathbb{C}}
\newcommand{\N}{\mathbb{N}}

\def\Re{\mathop{\mathrm{Re}}}
\def\Im{\mathop{\mathrm{Im}}}

\newcommand{\rmO}{\mathrm{O}}

\newcommand{\rmd}{\mathrm{d}}
\newcommand{\rme}{\mathrm{e}}
\newcommand{\rmi}{\mathrm{i}}

\newcommand{\Rg}{\mathrm{Rg}\,}

\renewcommand{\leq}{\leqslant}
\renewcommand{\geq}{\geqslant}

\newcommand{\GammaU}{(\nu_u^+-\nu_u^-)}
\newcommand{\GammaV}{(\nu_v^+-\nu_v^-)}

\def\XXint#1#2#3{{\setbox0=\hbox{$#1{#2#3}{\int}$}
     \vcenter{\hbox{$#2#3$}}\kern-.5\wd0}}

\newfam\bifam
\font\tenbi=cmmib10 scaled \magstep1 \font\sevenbi=cmmib10 at 11pt
\font\fivebi=cmmib10 at 6pt \textfont\bifam = \tenbi
\scriptfont\bifam = \sevenbi \scriptscriptfont\bifam= \fivebi

\begin{document}
\title{Criteria for pointwise growth and their role in invasion processes}


\author{Matt Holzer\footnote{current address: Department of Mathematical Sciences, George Mason University, Fairfax, VA 22030} \ and Arnd Scheel \\
University of Minnesota \\
School of Mathematics \\
127 Vincent Hall, 206 Church St SE \\
Minneapolis, MN 55455, USA}

\date{ \today}

\maketitle
\begin{abstract}
This article is concerned with pointwise growth and spreading speeds in systems of parabolic partial differential equations.  Several criteria exist for quantifying pointwise growth rates.  These include the location in the complex plane of singularities of the pointwise Green's function and pinched double roots of the dispersion relation.  The primary aim of this work is to establish some rigorous properties related to these criteria and the relationships between them.  In the process, we discover that these concepts are not equivalent and point to some interesting consequences for nonlinear front invasion problems.  Among the more striking is the fact that pointwise growth does not depend continuously on system parameters.  Other results include a determination of the circumstances under which pointwise growth on the real line implies pointwise growth on a semi-infinite interval.  As a final application, we consider invasion fronts in an infinite cylinder and show that the linear prediction always favors 
the formation of stripes in the leading edge.
\end{abstract}

\section{Introduction}

Invasion fronts play an important organizing role in spatially extended systems, with applications ranging from ecology to material science. They arise when a system is quenched into an unstable state and spatially localized fluctuations drive the system away into a more stable state. In an idealized situation, fluctuations are reduced to one small localized  perturbation of the initial state. Such an initial disturbance then grows and spreads spatially, leaving behind a new state of the system \cite{vS}. Beyond such an idealized scenario, one might expect that localization of fluctuations is quite unlikely in a large system. The mechanism of a spatially spreading disturbance is however still relevant, at least for the description of transients,  when initial disturbances are localized at several well separated locations in physical space \cite{reu2011}.  On the other hand, in particular for problems in ecology, unstable states prevail over large parts of space without disturbance because invasive 
species are simply absent in most of the domain, and spreading of the invasion is mediated by slow diffusive motion combined with exponential growth \cite{ecology}. In general, localization of disturbances can be achieved systematically when the system is quenched into an unstable state in a spatially uniformly progressing way, a scenario particularly relevant in a number of engineering applications \cite{goh}.

Conceptually, one is interested in two aspects of the invasion process: 
\begin{enumerate}
\item What is the invasion speed?
\item What is the state in the wake selected by the invasion process?
\end{enumerate}
The first question is most natural in ecological contexts while the second question occurs naturally in manufacturing and engineering \cite{engineering,water} applications. Both questions are clearly intimately related. One may envision, for instance, that different invasion speeds are associated with different possible states in the wake of the front and, in a simple scenario of almost linear superposition, the state in the wake of a primary invasion would be the fastest spreading state. 

The present work focuses mostly on the first question, while pointing out in some situations the intimate connection with the second aspect. 

In many simple, mostly scalar contexts, invasion speeds are well defined and can be characterized in various fashions, for instance using generalized eigenvalue problems or min-max principles \cite{ecology2,scalar,variational}. The key ingredient to almost all those characterizations is an order preservation property of the nonlinear evolution of the system. While very effective when available, such a property is intrinsically violated whenever we are interested in pattern-forming systems. 

In a somewhat less comprehensive and less rigorous fashion, spreading speeds have been known to be related to concepts of absolute and convective instability. Invasion speeds are characterized as critical states: an observer traveling at the spreading speed observes a marginally stable system \cite{dee,vS}. This characterization originates in the theory of absolute and convective instability, motivated originally by studies of plasma instabilities \cite{bri} with many applications in fluid dynamics \cite{bers}. Without striving to give a comprehensive (or even adequate) review of the relevant literature, we will pursue this approach in a systematic fashion. Trying to press some folklore observations into precise lemmas, we uncover a number of apparently unknown (or at least under-appreciated) aspects of convective and absolute instabilities, which directly impact the characterization of spreading speeds. 

As a general rule, the analysis here is linear, but intrinsically motivated by the desire to derive criteria and consequences for nonlinear invasion processes. Beyond providing a precise language for the mathematically inclined reader interested in this approach to invasion problems, we point to several interesting phenomena that deserve further exploration.  In particular, we highlight two main results of this work:

\paragraph{Multiple Spreading Modes.} We show that a number of intriguing subtleties are associated with multiple, degenerate spreading modes. We show that in this case growth modes and spreading speeds lack continuity properties with respect to system parameters and point to consequences for nonlinear invasion problems. 

\paragraph{Multi-Dimensional Spreading forms Stripes.}
We prove a fundamental result on multi-dimensional invasion processes which states, loosely speaking, that linearly determined multi-dimensional pattern-forming invasion processes \emph{always} select stripes, or one-dimensional patterns, in the leading edge. More complex patterns such as squares or hexagons are always consequences of secondary invasion processes.

\paragraph{Outline.} This paper is organized as follows. We will characterize pointwise growth rates through pointwise Green's functions in Section \ref{s:1}, thus distinguishing between convective and absolute instabilities in a quantitative and systematic fashion. In Section \ref{s:2a}, we consider the positive half line and the influence of boundary conditions on the pointwise growth rates.  In Section \ref{s:alg}, we recall the classical characterization of pointwise growth rates via pinched double roots. We show that pinched double roots may overestimate pointwise growth rates but are generically equivalent to singularities of the Green's function. We discuss more generally properties of both concepts in Section \ref{s:prop}.  In Section \ref{s:sp}, we introduce comoving frames and spreading speeds. Section \ref{s:spmult} contains our main result on multi-dimensional spreading behavior. Finally, we discuss several extensions in Section \ref{s:dis} such as nonlinear problems, problems with periodic 
coefficients, and localized spatial inhomogeneities. 

\begin{Acknowledgment}
The authors acknowledge partial support through NSF (DMS-1004517  (MH), DMS-0806614(AS), DMS-1311740(AS)).  This research was initiated during an NSF-Sponsored REU program in the summer of 2012 \cite{reu2012}.  We are grateful to Koushiki Bose, Tyler Cox, Stefano Silvestri and Patrick Varin for working out some of the examples in this article.  We also thank Ryan Goh for many stimulating discussions related to the material in Sections \ref{s:alg} and \ref{s:prop}.
\end{Acknowledgment}

\section{Pointwise Growth Rates --- The Green's Function and Pointwise Projections}
\label{s:1}
In this section, we study pointwise growth in terms of properties of the convolution kernel of the resolvent. We focus on parabolic systems, Section \ref{s:1.1}, discuss differences between pointwise stability and stability in norm, Section \ref{s:1.2}, and give a general, geometric characterization of pointwise growth in Section \ref{s:1.3}. We conclude with a number of examples that will also be relevant later, Section \ref{sec:examples}.

\subsection{Setup}\label{s:1.1}
In this section, we will review the notion of convective and absolute instabilities in dissipative systems.  In particular, we study instabilities related to invasion phenomena in reaction-diffusion systems. First consider a general system of parabolic equations 
\begin{equation}\label{e:0}
u_t = A(\nabla_x)u, \quad u\in\R^N, \ x\in\R^n,
\end{equation}
where $A$ is a constant coefficient elliptic operator of order $2m$.  That is, $A$ is a vector-valued polynomial of order $2m$ such that the $2m$'th order coefficients satisfy strict ellipticity. More explicitly, we write
\[
A(p)=\sum_{j=0}^{2m} A_j(p),\quad A_j(\nabla_x)=\sum_{|\ell|=j}A_{j,\ell}\partial_x^\ell,
\]
with multi-index notation $\ell \in \N^n$ so that $|\ell| = \sum \ell_i$ and $\partial_x^\ell = \partial_{x_1}^{\ell_1} \ldots \partial_{x_n}^{\ell_n}$. We then require that the $N \times N$-matrices $A_{2m,\ell}$ satisfy strict ellipticity.  Equivalently, there exists some $\delta > 0$ such that
\begin{equation}\label{e:cond}
\sum_{|\ell|=2m,i,j} A^{ij}_{2m,\ell} (-1)^{m+1} k^\ell u_i u_j \geq \delta |k|^{2m} |u|^2
\end{equation}
for all $k \in \R^n$ and $u \in \R^N$.
Our main interest is in ``generic'' compactly supported (smooth) initial conditions $u_0(x)$ and their behavior as $t\to\infty$. 

\subsection{Stability and Instability --- $L^2$ and Exponential Weights}\label{s:1.2}

The  spectrum of $A(\nabla_x)$ in translation-invariant spaces such as $L^2(\R,\R^N)$ consists of continuous essential spectrum only, that is, $A-\lambda$ is not Fredholm with index 0 when it is not invertible. The spectrum can be determined from the dispersion relation
\[
d(\lambda,\nu) := \mathrm{det}\,(A(\nu)-\lambda)
\]
as 
\[
\sigma (A)=\{\lambda; d(\lambda,\nu)=0 \mbox{ for some }\nu\in\C^n \mbox{ with } \Re\nu=0\}.
\]
This can be readily established in $L^2(\R,\R^N)$ using a Fourier transform, but it also holds in most other translation-invariant norms such as $BC^0_\mathrm{unif}$ and $L^p$. 

We say $A$ is stable if $\Re\sigma(A)<0$ and $A$ is unstable if $\sigma(A)\not\subset \{\Re\lambda\leq 0\}$. 

On the other hand, choosing exponentially weighted norms,
\begin{equation}
\|u\|_{L^2_\eta}=\|u(\cdot)\rme^{(-\eta,\cdot)}\|_{L^2}
\end{equation}
for some weight-vector $\eta$, one finds 
\[
\sigma_\eta(A)=\{\lambda; d(\lambda,\nu)=0 \mbox{ for some }\nu\in\C^n \mbox{ with } \Re\nu=\eta\}.
\]
This can be readily established using the isomorphism 
\[
\iota_\eta:L^2_\eta\to L^2, \quad u(\cdot)\mapsto u(\cdot)\rme^{(-\eta,\cdot)},
\]
which conjugates $A(\nabla_x)$ and $A(\nabla_x + \eta)$. 

Our interest here is in a slightly different notion of stability, where one poses the system on $\R^n$ but measures stability or instability only in bounded regions $\Omega$ of $\R^n$. We will see that stability and instability properties do not depend on the particular choice of the window of observation $\Omega$ and we refer to this type of stability property as \emph{pointwise stability} or \emph{pointwise instability}. To be precise, consider the solution $G(t,x)u_0$ with Dirac initial condition $u_0 \delta(x)$, $u_0\in\R^N$. 
We say $A$ is pointwise stable if $\sup_{|x|\leq 1}|G(t,x)|\leq C\rme^{-\delta t}$ for some $C,\delta>0$. We say $A$ is pointwise unstable if $\sup_{|x|\leq 1}|G(t,x)|\geq C\rme^{\delta t}$ for some $C,\delta>0$.

Using parabolic regularity, one readily finds 
\begin{itemize}
\item pointwise instability $\Longrightarrow$ $L^2$-instability
\item $L^2$-stability $\Longrightarrow$ pointwise stability
\end{itemize}

The intermediate regime of pointwise stability and $L^2$-instability is often referred to as \emph{convective instability}, while pointwise instability is also often called \emph{absolute instability}. 

From the estimates on the essential spectrum, we  obtain that 
\[
\|u(t,\cdot)\|_{L^2(\R^n)}\leq C\rme^{\delta t}\|u(0,\cdot)\|_{L^2(\R^n)},
\]
where $\delta > \max\{\Re\sigma(A)\}$ and $C = C(\delta)>0$. This estimate is in fact sharp by spectral mapping theorems which hold for sectorial operators of the form considered here; see for instance \cite{Lun}. 

On the other hand, for $u(0,\cdot)$ compactly supported on $\Omega$, the observation on exponential weights shows that 
\begin{equation}\label{e:ref}
\|u(t,\cdot)\|_{L^2(\Omega)}\leq C\rme^{\delta t}\|u(0,\cdot)\|_{L^2(\R^n)},
\end{equation}
with  any
\[
\delta>\inf_\eta \max\{\Re\sigma_\eta\,(A)\}.
\]
In general, this estimate is not sharp in a fundamental way. For instance, the best choice of $\eta$ might depend on the wavenumber $k$. In a more subtle way, unstable absolute spectrum \cite{SSabs} may give rise to instabilities in \emph{any} exponential weight while not generating pointwise instabilities in the above sense. We will discuss this particular discrepancy and resulting complications later when pointing out differences between pinched double roots and pointwise growth rates.

We will concern ourselves with the one-dimensional case where $n = 1$ and $x \in \R$. We discuss multi-dimensional instabilities in Section \ref{s:spmult}. 

Consider the one-dimensional restriction of (\ref{e:0}), 
\begin{equation}\label{e:01}
u_t=A(\partial_x)u, \quad u\in\R^N,
\end{equation}
where 
\[
A(\partial_x)=\sum_{j=0}^{2m} A_j\partial_x^j,
\]
and 
\begin{equation}\label{e:ell}
\Re\sigma((-1)^mA_{2m})<0.
\end{equation}
Note that this condition is seen to be equivalent to (\ref{e:cond}) for a suitable choice of coordinates in $\R^N$. 

\subsection{Pointwise Growth -- The Analytic Extension of the Green's Function}\label{s:1.3}

In order to solve (\ref{e:0}), we use the Laplace transform to write
\begin{equation}\label{e:1}
u(t,x)=\frac{-1}{2\pi\rmi}\int_\Gamma \rme^{\lambda t}(A-\lambda)^{-1}u(0,x)\rmd\lambda.
\end{equation}
The contour $\Gamma$ is initially chosen to be sectorial and to the right of the spectrum of $A$.  Further, the resolvent is given as a convolution with the Green's function
\[
[(A-\lambda)^{-1}f](x)=\int_yG_\lambda(x-y)f(y)\rmd y.
\]
The Green's function in turn can be found via the inverse Fourier transform $\mathcal{F}$ of the Fourier conjugate operator $A(\rmi k)$,
\begin{equation}\label{e:f}
G_\lambda(\xi)=\mathcal{F}\left[(A(\rmi k)-\lambda)^{-1}\right].
\end{equation}
When $u(0,x)$ has compact support we can evaluate (\ref{e:1}) pointwise, 
\begin{equation}\label{e:2}
u(t,x)=\frac{-1}{2\pi\rmi}\int_{\Gamma'} \rme^{\lambda t}\int_y G_\lambda(x-y)u(0,y)\rmd  y\rmd\lambda.
\end{equation}
The contour $\Gamma'$ can be any contour with the same asymptotics as $\Gamma$ when $\Re\lambda\to -\infty$ so that $G_\lambda(\xi)$ is analytic in the region bounded by $\Gamma$ and $\Gamma'$. In particular, we can choose $\Re\Gamma'<0$ provided that $G_\lambda(\xi)$ does not possess singularities in $\Re\lambda\geq 0$.  This is much weaker than the typical condition $G_\lambda\in L^1$ for the boundedness of the resolvent in $L^2$.  Similar to the discussion on exponential weights in (\ref{e:ref}), one could for instance compute the Fourier transform (\ref{e:f}) by shifting the contour of integration $k\in\R$ in the complex plane to $k\in\R+\rmi\eta$. 

We can also construct $G_\lambda(\xi)$ directly using the reformulation of the resolvent equation $(A-\lambda)u=f$ as a first-order constant-coefficient ODE
\begin{equation}\label{e:1st}
U_x=M_\lambda U+F,\quad U=(u,\frac{\rmd}{\rmd x} u,\ldots,\frac{\rmd^{2m-1}}{\rmd x^{2m-1}} u),
\end{equation}
for $U\in\mathbb{R}^{2mN}$.  Denote by $T_\lambda(\xi)$ the Greens function to this first order equation. Then 
\begin{equation}\label{e:1n}
G_\lambda(\xi)=P_1T_\lambda(\xi)Q_1A_{2m}^{-1}
,
\end{equation}
where  $P_1$ is the projection on the first component and $Q_1$ is the embedding into the last component.  That is, $P_1U=u$ and $Q_1f=(0,\ldots,0,f)$. The first-order Green's function is determined by the decomposition into stable and unstable subspaces $E^\mathrm{s}_\lambda$ and $E^\mathrm{u}_\lambda$, with associated spectral projections $P^\mathrm{s}_\lambda=\mathrm{id}-P^\mathrm{u}_\lambda$. Indeed, with $\Phi_\lambda(x)$ denoting the (linear, $F=0$) flow to (\ref{e:1st}), 
\begin{equation}\label{e:ge}
T_\lambda(\xi) = \begin{cases}
\Phi_\lambda(\xi)P^\mathrm{s}_\lambda , & \xi > 0, \\
-\Phi_\lambda(\xi)P^\mathrm{u}_\lambda , & \xi < 0.
\end{cases}
\end{equation}
Note that the definition of $E^\mathrm{s/u}_\lambda$ is unambiguous for $\lambda$ large due to ellipticity of $A(\partial_x)$.  Moreover, $\dim\,(E^\mathrm{s/u}_\lambda)=mN$. To see this, note that for large $\lambda$, eigenvalues of $M_\lambda$ with multiplicities are (to leading order) roots of $A_{2m}\nu^{2m}-\lambda=0$.  This implies $|\lambda|\sim |\nu|^{1/2m}$ and $\Re \nu\neq 0$ by ellipticity of $A_{2m}$, (\ref{e:ell}). Since, to leading order, roots $\nu$ come in pairs $\nu$ and $-\nu$, we have that $\dim\,(E^\mathrm{s/u}_\lambda)=mN$.

We are interested in the extension of the Green's function and the subspaces $E^\mathrm{s/u}_\lambda$ as functions of $\lambda$ and possible singularities in the form of poles or branch points.  We first demonstrate that singularities of these objects occur simultaneously. 

\begin{Lemma}\label{l:1st-2nd}
The following regions coincide:
\begin{itemize}
\item $D_G=\{\lambda\in\C| G_\lambda \mbox{ is analytic in }\lambda\}$;
\item  $D_T=\{\lambda\in\C| T_\lambda \mbox{ is analytic in }\lambda\}$;
\item  $D_P=\{\lambda\in\C| P^\mathrm{s}_\lambda \mbox{ is analytic in }\lambda\}$.
\end{itemize}
\end{Lemma}
\begin{Proof}
Since $M_\lambda$ is analytic, $\Phi_\lambda$ is analytic and (\ref{e:ge}) shows that $D_T=D_P$. Also, (\ref{e:1n}) shows that $D_T\subset D_G$.  It remains to show that analyticity of $G_\lambda$ implies analyticity of $T_\lambda$. We will therefore construct $T_\lambda$ explicitly from $G_\lambda$ and its derivatives. 

We need to solve  $U_x=M_\lambda U + F$, which we write more explicitly as 
\begin{align}\label{rhs}
u_i'&= u_{i+1}+F_i, \quad \quad i=1,\ldots, 2m-1,\nonumber\\
u_{2m}'&=-A_{2m}^{-1}\left(\sum_{i=0}^{2m-1}A_iu_{i+1}+\lambda u_1\right)+F_{2m}.
\end{align}
We change variables for $i = 1, \ldots, 2m-1$, setting  $ \tilde{u}_{i+1}=u_{i+1}+\sum_{j=1}^{i} F_j^{(i-j)}$, where we write $F^{(k)}:=\frac{\rmd}{\rmd x}F$. 
Note that $u_1$ is left unchanged, and hence $\tilde{u}_1=u_1$.
Then the system (\ref{rhs}) becomes
\begin{align}\label{tilde}
\tilde{u}_i'&= \tilde{u}_{i+1}, \quad \quad i=1,\ldots, 2m-1,\\
\tilde{u}_{2m}'&=-A_{2m}^{-1} \left(\sum_{i=0}^{2m-1}A_i\tilde{u}_{i+1}
-\sum_{j=1}^{2m-1}A_j\sum_{k=1}^{j}F_k^{(j-k)}+\lambda u_1\right)+\sum_{j=1}^{2m-1}F_j^{(2m-j)}+F_{2m}.\nonumber
\end{align}
\\
We may then solve for $u_1$ by noting that
\[
u_1= \tilde{u}_1=G_\lambda*\left(A_{2m}F_{2m}+A_{2m}\sum_{j=1}^{2m-1}F_j^{(2m-j)}+\sum_{j=1}^{2m-1}A_j\sum_{k=1}^{j}F_k^{(j-k)}\right).
\]
The values for $u_i$, $i>1$,  are obtained in a similar manner using (\ref{tilde}) and the change of coordinates relating $u$ to $\tilde{u}$.  In order to relate the two Green's functions $T_\lambda$ and $G_\lambda$, we must write these expressions as convolutions against the inhomogeneous terms $F_i(x)$.  Since derivatives of these functions exist in the definition of $u_1$, we must integrate by parts and this introduces derivatives of the pointwise Green's function $G_\lambda$ into the expressions.  Any derivative of $G_\lambda$ that has order greater than $2m-1$ can be eliminated by using the defining property of $G_\lambda$, that is, \[ A_{2m}G_\lambda^{(2m)}+\sum_{j=0}^{2m-1}A_j G_\lambda^{(j)}=\delta(x),\]
where $\delta(x)$ is the Dirac delta function and superscripts refer to derivatives.  In this fashion, we are able to compute $T_\lambda$ as a function of $G_\lambda$ and its derivatives.  For example, the first $N$ rows of $T_\lambda$ are
\[ t_{1,j}=\sum_{i=j}^{2m-1}G_\lambda^{(i-j)}A_i+G_\lambda^{(2m-j)}A_{2m}.\]

Further entries can be computed by a straightforward, albeit tedious, calculation.  

All that remains is to show that the derivatives of $G_\lambda$ involved in the definition of $T_\lambda$ are analytic whenever $G_\lambda$ is analytic.  Let $J$ be an interval $J\subset \mathbb{R}$.  Define the operator
\begin{eqnarray*}
\mathcal{H}_\lambda^\delta &:& (C^{2m}(J))^N\to (C^0(J))^N\times \mathbb{R}^{2mN} \\
&\ & u(x)\mapsto \left(A(\partial_x)u-\lambda u, u(0),u(\delta),\dots,u((2m-1)\delta)\right).
\end{eqnarray*}
This operator is conjugate to the operator $\tilde{\mathcal{H}}_\lambda^\delta$, which maps $u(x)$ to
\[\left(A(\partial_x)u-\lambda u, u(0),\Delta_\delta u(0),\Delta_\delta^2 u(0),\dots ,\Delta_\delta^{(2m-1)}u(0))\right),\]
where $\Delta_\delta$ is the discrete forward difference operator.  For $\delta>0$ and sufficiently small, $\tilde{\mathcal{H}}_\lambda^\delta$ is, in turn,  $\delta$-close to the operator $\tilde{\mathcal{H}}_\lambda$ which maps $u(x)$ to $\left(A(\partial_x)u-\lambda u, u(0),u'(0),\dots,u^{(2m-1)}(0))\right)$.  Standard existence and uniqueness implies that $\tilde{\mathcal{H}}_\lambda$ is invertible and consequently both $\tilde{\mathcal{H}}_\lambda^\delta$ and $\mathcal{H}_\lambda^\delta$ are invertible for $\delta>0$ sufficiently small.  We then have \[
G_\lambda(x)=(\mathcal{H}_\lambda^\delta)^{-1}(0,G_\lambda(0),\dots,G_\lambda((2m-1)\delta)).\]
More importantly, $G_\lambda(x)$ is an analytic function of $\lambda$ as an element of $C^{2m}(J)^N$.  Then, for fixed $x\in J$ derivatives up to order $2m$ can be computed and are analytic.  This establishes the analyticity of $T_\lambda(x)$ given the analyticity of $G_\lambda(x)$ and completes the proof of Lemma~\ref{l:1st-2nd}.

%
\end{Proof}

The above discussion roughly guarantees  upper bounds for pointwise growth in terms of singularities of $P^\mathrm{s}_\lambda$. This motivates the following definition. 

\begin{Definition}[Pointwise Growth Modes and Rates]\label{d:pgr}
We say $\lambda_*$ is a \emph{pointwise growth mode} (PGM) if $P_\lambda^\mathrm{s}$ is not analytic in $\lambda$ at $\lambda_*$. 
The \emph{pointwise growth rate} (PGR) is defined as the maximal real part of a pointwise growth mode, or,
\[
\inf \{\rho;\, P_\lambda^\mathrm{s}\mbox{ is analytic in }\Re\lambda>\rho\}.
\]
\end{Definition}

The following result shows that, in an appropriate sense, pointwise growth modes give sharp characterizations of pointwise growth. 

\begin{Corollary}[PGMs are Sharp]\label{c:lb}
The pointwise growth rate $\rho$ defines generic pointwise growth of the solutions as follows. 
\begin{itemize}
\item \emph{Upper bounds:} For  any $\rho'>\rho$, any compactly supported initial condition $u_0(x)$, and any fixed interval $(-L,L)$, we have for the solution $u(t,x)$ with initial condition $u(t,x)=u_0(x)$ 
\begin{equation}\label{e:ub}
\limsup_{t\to\infty}\sup_{x\in (-L,L)} |u(t,x)|\rme^{-\rho't}=0.
\end{equation}
\item \emph{Lower bounds:} For any $\rho'<\rho$, there exists $v\in\C^N$ and $L>0$ so that the solution $u(t,x)$  with initial condition $u(0,x)=v\delta(x)$ satisfies 
\begin{equation}\label{e:lb}
\limsup_{t\to\infty}\sup_{x\in (-L,L)} |u(t,x)|\rme^{-\rho't}=\infty.
\end{equation}
\end{itemize}
\end{Corollary}
\begin{Proof}
The upper bounds were established in Lemma \ref{l:1st-2nd}. Lower bounds can be obtained indirectly. Suppose that we had upper bounds of the form (\ref{e:ub}) with $\rho'<\rho$ for all $L>0$. Considering the solution with initial data $\delta(x)$, we find exponential decay for the fundamental solution $|S(t,x)|\leq C\rme^{\rho' t}$. Taking the Laplace transform of $S(t,x)$ yields the Green's function $G_\lambda(x)$ which is analytic in $\Re\lambda>\rho'$,  contradicting the assumption.
\end{Proof}

%
%

\begin{Remark}[Convective and Absolute Instabilities]
For an $L^2$-unstable system, we find a trichotomy:
\begin{itemize}
\item if the pointwise growth rate is negative, we say that the instability is \emph{convective};
\item if the pointwise growth rate is positive, we say that the instability is \emph{absolute};
\item if the pointwise growth rate is zero, we say that the instability is (pointwise) \emph{marginal};
\end{itemize}
\end{Remark}

\subsection{Examples}\label{sec:examples}
We will collect a set of examples that will serve as illustrations for some of the more subtle effects that we shall discuss later on. In these examples, we compute the Green's function and $P^\mathrm{s}_\lambda$ and point out the correspondence between singularities of the Green's function and singularities of $P^\mathrm{s}_\lambda$ from Lemma  \ref{l:1st-2nd}.

\paragraph{Convection-Diffusion.}
The first example that we consider is the scalar convection-diffusion equation,
\begin{equation}
u_t=u_{xx}+u_x.\label{convection-diffusion}
\end{equation}
We follow the procedure outlined above.  A Laplace transform in time converts the PDE into a system of first order ordinary differential equations which reads, adopting the notation from (\ref{e:1st}),
\[
U_x = M_\lambda U = \begin{pmatrix}
0 & 1 \\
\lambda & -1
\end{pmatrix}
U.
\]
The fundamental matrix \( \Phi_{\lambda}(\xi) \) associated to this system can be computed from the eigenvalues and eigenvectors of $M_\lambda$,
\[ \nu_\pm(\lambda)=-\frac{1}{2}\pm\frac{1}{2}\sqrt{1+4\lambda}.\]
Let $L$ be the matrix whose rows are the eigenvectors corresponding to the stable left eigenvalues, and let $R$ be the matrix whose columns are the eigenvectors corresponding to the stable right eigenvalues.  Then \( P_{\lambda}^s = R (LR)^{-1}L \).  That is,
\begin{align}
P_{\lambda}^s &=\frac{1}{\nu_+(\lambda)-\nu_-(\lambda)}  \begin{pmatrix}
\nu_+(\lambda) & -1 \\
\nu_-(\lambda)\nu_+(\lambda) & -\nu_-(\lambda)\end{pmatrix}\nonumber\\
&= \frac{1}{\sqrt{1+4\lambda}}\begin{pmatrix} -\frac{1}{2}+\frac{1}{2}\sqrt{1+4\lambda} & -1 \\
-\lambda & \frac{1}{2}+\frac{1}{2}\sqrt{1+4\lambda}
\end{pmatrix}.
\label{ps-conv-diff}
\end{align}
We note that the eigenvalues $\nu_\pm(\lambda)$ are analytic and distinct away from $\lambda=-\frac{1}{4}$.  This implies that $P^\mathrm{s}_\lambda$ is analytic away from $\lambda=-\frac{1}{4}$ as well. The Green's function for the first order equation, $T_\lambda$, is given through (\ref{e:ge}), and, since there is only one stable and unstable eigenvalue, we find
\begin{equation}
T_\lambda(\xi) = \begin{cases}
e^{\nu_-(\lambda)\xi}P^\mathrm{s}_\lambda , & \xi > 0, \\
-e^{\nu_+(\lambda)\xi}P^\mathrm{u}_\lambda , & \xi < 0.
\end{cases}
\end{equation}
As noted in Lemma \ref{l:1st-2nd}, $T_\lambda(\xi)$ possesses the same domain of analyticity as $P_\lambda$. 

Finally, using the relation \( G_{\lambda}(\xi) = P_1 T_{\lambda}(\xi) Q_1 \), we find that the Green's function \( G_{\lambda}(\xi) \) is given by
\begin{equation*}
G_{\lambda}(\xi) = \begin{cases}
- \frac{1}{\sqrt{1+4\lambda}}e^{\nu_-(\lambda)\xi}, & \xi > 0,  \\
- \frac{1}{\sqrt{1+4\lambda}}e^{\nu_+(\lambda)\xi}, & \xi < 0. 
\end{cases}
\end{equation*}
In agreement with Lemma \ref{l:1st-2nd}, $G_\lambda$ inherits the analyticity properties of $T_\lambda$: it is analytic except for a singularity at $\lambda=-\frac{1}{4}$.

%
%
%
%
%
%
%

\paragraph{Cahn-Hilliard.}
Consider the following parabolic partial differential equation, 
\begin{equation}
u_t=-u_{xxxx}-\mu u_{xx}.\label{cahn-hilliard}
\end{equation}
This equation is encountered, for example, as the linearization of the Cahn-Hilliard equation at a homogeneous steady state.  If $\mu<0$ the steady state is stable and if $\mu>0$ it is unstable with respect to long wavelength perturbations.  We will compute the pointwise Green's function and determine its singularities.  After a Laplace transform in time, the system is a fourth order ordinary differential equation with four eigenvalues,
\[ \pm\sqrt{-\frac{\mu}{2}\pm\frac{1}{2}\sqrt{\mu^2-4\lambda}}.\]
We define the principle value of the square root to lie in the upper half of the complex plane.  Then the stable eigenvalues are 
\begin{equation} -\sqrt{-\frac{\mu}{2}+\frac{1}{2}\sqrt{\mu^2-4\lambda}}, \ \ \text{and} \ \ +\sqrt{-\frac{\mu}{2}-\frac{1}{2}\sqrt{\mu^2-4\lambda}}.\label{eq:stableCH} \end{equation}
For any eigenvalue $\nu$, the stable right eigenvector is $(1,\nu,\nu^2,\nu^3)^T$ and left eigenvector is $(\nu^3+\nu\mu,\nu^2+\mu,\nu,1)$.  Using these vectors and the formula $P_\lambda=R(LR)^{-1}L$ the projection onto this eigenspace is
\begin{equation}
P_\lambda^\nu=\frac{1}{4\nu^3+2\mu\nu}\left( \begin{array}{cccc} 
\nu^3+\nu\mu & \nu^2+\mu & \nu & 1 \\
\nu^4+\nu^2\mu & \nu^3+\mu\nu & \nu^2 & \nu \\
\nu^5+\nu^3\mu & \nu^4+\mu\nu^2 & \nu^3 & \nu^2 \\
\nu^6+\nu^4\mu & \nu^5+\mu\nu^3 & \nu^4 & \nu^3 \end{array}\right).
\end{equation}
It is then straightforward to write down the stable projection as the sum of the stable projections associated to the two stable eigenvalues (\ref{eq:stableCH}).  Explicit formulas for the first order Green's function are rather involved, so we skip straight to the pointwise Green's function,
\begin{equation}
G_\lambda(\xi)= \left\{ \begin{array}{c}
-\frac{\sqrt{-\frac{\mu}{2}+\frac{1}{2}\sqrt{\mu^2-4\lambda}}}{\sqrt{4\lambda}\sqrt{\mu^2-4\lambda}}e^{\xi\sqrt{-\frac{\mu}{2}-\frac{1}{2}\sqrt{\mu^2-4\lambda}}} -\frac{\sqrt{-\frac{\mu}{2}-\frac{1}{2}\sqrt{\mu^2-4\lambda}}}{\sqrt{4\lambda}\sqrt{\mu^2-4\lambda}}e^{-\xi\sqrt{-\frac{\mu}{2}+\frac{1}{2}\sqrt{\mu^2-4\lambda}}},\ \ \xi>0 \\
-\frac{\sqrt{-\frac{\mu}{2}+\frac{1}{2}\sqrt{\mu^2-4\lambda}}}{\sqrt{4\lambda}\sqrt{\mu^2-4\lambda}}e^{-\xi\sqrt{-\frac{\mu}{2}-\frac{1}{2}\sqrt{\mu^2-4\lambda}}} -\frac{\sqrt{-\frac{\mu}{2}-\frac{1}{2}\sqrt{\mu^2-4\lambda}}}{\sqrt{4\lambda}\sqrt{\mu^2-4\lambda}}e^{\xi\sqrt{-\frac{\mu}{2}+\frac{1}{2}\sqrt{\mu^2-4\lambda}}},\ \ \xi<0. \end{array}\right. \label{eq:ptGCH}
\end{equation}

Observe that  \( G_{\lambda}(\xi) \) has singularities when $\lambda=0$ and when $\lambda=\frac{\mu^2}{4}$.  We will focus on the singularity at $\lambda=\frac{\mu^2}{4}$.  We will see that the nature of this singularity depends on the sign of $\mu$.  When $\mu<0$, the singularity is removable.  This can be seen as follows.  Consider a fixed value of $\xi>0$ and let $\lambda\to\frac{\mu^2}{4}$ with $\lambda>\frac{\mu^2}{4}$.   Since $\mu<0$, the factors  $-\mu\pm\sqrt{\mu^2-4\lambda}$ inside the roots in (\ref{eq:stableCH}) converge to a common value on the positive real axis.  Due to our choice of the principle value of the root, we then observe that the roots  $\sqrt{-\frac{\mu}{2}\pm\frac{1}{2}\sqrt{\mu^2-4\lambda}}$ lie in the upper half plane and converge as $\lambda\to\frac{\mu^2}{4}$ to the values  $\pm\sqrt{-\mu/2}$.  Consequently, the pointwise Green's function has a finite limit as $\lambda\to\frac{\mu^2}{4}$ and the singularity is removable.

This should be contrasted with the case when $\mu>0$.  Now the two factors  $-\mu\pm\sqrt{\mu^2-4\lambda}$ converge to a common point on the negative real axis as $\lambda\to\frac{\mu^2}{4}$. Upon taking the root, the two factors $\sqrt{-\mu\pm\sqrt{\mu^2-4\lambda}}$ converge to the purely imaginary value $i\sqrt{\mu/2}$ in the limit.  As a result, the cancellation mechanism that was at play in the case of $\mu<0$ no longer holds and the singularity is a pole for $\mu>0$.



\paragraph{Counter-Propagating Waves.}
The following example illustrates an important subtlety. The subspaces $E^\mathrm{s/u}_\lambda$ are of course analytic as elements of the Grassmannian $\mathrm{Gr}_\C(2m,m)$ whenever $P^\mathrm{s}_\lambda$ is analytic. However, the converse is not true since $E^\mathrm{s}_\lambda$ and $E^\mathrm{u}_\lambda$ may intersect.  We consider the following system,
\begin{align}
u_t&=u_{xx}+u_x\nonumber\\
v_t&=v_{xx}-v_x+\mu u\label{e:cp}
\end{align}
As in the previous examples, we begin by transforming the system of second order equations into a system of first order equations.  We use the standard ordering $(u,u_x,v,v_x)$ in contrast to (\ref{e:1st}) and obtain
\[
U_x = M_\lambda U = \begin{pmatrix}
0 & 1 & 0 & 0 \\
\lambda & -1 & 0 & 0 \\
0 & 0 & 0 & 1 \\
- \mu & 0 & \lambda & 1 \\
\end{pmatrix}
U.
\]
The eigenvalues of this system are determined by the eigenvalues of the blocks corresponding to the $u$ and $v$ systems in isolation.  There, we have
\[
\nu_u^\pm=-\frac{1}{2}\pm\frac{1}{2}\sqrt{1+4\lambda}, \ \nu_v^\pm=\frac{1}{2}\pm\frac{1}{2}\sqrt{1+4\lambda}.\]
The stable and unstable eigenvalues and eigenvectors give rise to the eigenspaces, 
\begin{equation}\label{e:esu}
E^\mathrm{s}_\lambda=\mathrm{Span}\left(\begin{array}{cc} 1& 0\\ \nu_u^- &0\\ \frac{\mu}{2\nu_u^-} &1\\ \frac{\mu}{2} &\nu_v^-\end{array}\right) \quad \text{and} \quad
E^\mathrm{u}_\lambda=\mathrm{Span}\left(\begin{array}{cc} 1 & 0\\ \nu_u^+ &0\\ \frac{\mu}{2\nu_u^+}&1 \\\frac{\mu}{2} &\nu_v^+\end{array}\right).
\end{equation}

Following the general procedure outlined above, we obtain the stable projection as well as the Green's function.  The stable projection is,
\[ P_\lambda^s=\left(\begin{array}{cc} P_{\lambda,u}^s & 0 \\ P_{\lambda,u\to v}^s & P_{\lambda,v}^s \end{array}\right),\]
where the diagonal elements are the projections for the $u$ and $v$ sub-systems in isolation and are similar to the stable projection (\ref{ps-conv-diff}).  The sub-matrix $P_{\lambda,u\to v}^s$ describes the effect of the coupling and is given by 
\[ P_{\lambda, u\to v}^s=\mu \left(\begin{array}{cc} \frac{1+\nu_v^-}{2\nu_v^-\GammaV}+\frac{\nu_u^+}{2\nu_u^-\GammaU} & \frac{-1}{2\nu_u^-\GammaU}+\frac{1}{2\nu_v^-\GammaV} \\
\frac{1+\nu_v^-}{2\GammaV}+\frac{\nu_u^+}{2\GammaU} & \frac{-1}{2\GammaU}+\frac{1}{2\GammaV}\end{array}\right).\]
The explicit formula in terms of $\lambda$ is,
\[ P_{\lambda,u\to v}^s=\mu \left(\begin{array}{cc} \frac{-1}{2\lambda\sqrt{1+4\lambda}} & \frac{-1}{2\lambda\sqrt{1+4\lambda}} \\
\frac{1}{2\sqrt{1+4\lambda}} & 0 \end{array}\right).\]
Note that all components of the stable projection have a singularity at $\lambda=-\frac{1}{4}$.  Only the projection matrix $P_{\lambda,u\to v}^s$ has a second singularity at $\lambda=0$.  This singularity is a pole and, in this respect, is fundamentally distinct from the singularity at $\lambda=-\frac{1}{4}$. 


With the projections determined, we can compute the pointwise Green's function.  We have,
\begin{equation*}
G_{\lambda}(\xi) = \begin{cases}
\mathcal{A}_1, & \xi > 0,  \\
\mathcal{A}_2, & \xi < 0,
\end{cases}
\end{equation*}
where
\begin{align*}
\mathcal{A}_1 &= \frac{e^{\nu_u^-\xi}}{\sqrt{1+4\lambda}}\left(\begin{array}{cc}-1 & 0 \\ \frac{\mu(1-\sqrt{1+4\lambda})}{4\lambda} & 0\end{array}\right)
+\frac{e^{\nu_v^-\xi}}{\sqrt{1+4\lambda}}\left(\begin{array}{cc}0 & 0 
\\ \frac{\mu(1+\sqrt{1+4\lambda})}{4\lambda}  & -1\end{array}\right) 
\\
\mathcal{A}_2 &= \frac{e^{\nu_u^+\xi}}{\sqrt{1+4\lambda}}\left(\begin{array}{cc}1 & 0 \\ \frac{-\mu(1+\sqrt{1+4\lambda})}{4\lambda} & 0\end{array}\right)
+\frac{e^{\nu_v^+\xi}}{\sqrt{1+4\lambda}}\left(\begin{array}{cc}0 & 0 
\\ \frac{-\mu(1-\sqrt{1+4\lambda})}{4\lambda}  & 1\end{array}\right) 
\end{align*}
For $\mu\neq 0$, $P^\mathrm{s}_\lambda$ and $G_\lambda$ both possess a singularities at $\lambda=0$ and $\lambda=-\frac{1}{4}$.  For $\mu=0$, the singularity at $\lambda=0$ disappears.  We also remark that the subspaces $E^\mathrm{s}$ and $E^\mathrm{u}$ are analytic in $\lambda$ for all $\lambda\neq-\frac{1}{4}$.  At $\lambda=0$ they intersect non-trivially, but remain analytic.  

\begin{Remark}[Upper Semi-Continuity of PGR]\label{r:usc}
In all the previous example, one can verify explicitly that pointwise growth modes depend continuously on system parameters. In the present example of counter-propagating waves, however, a pointwise growth mode disappears at the specific value $\mu=0$, so that the pointwise growth rate is only upper semi-continuous. We will generalize this observation later, Lemma \ref{l:jump}.
\end{Remark}

\begin{Remark}[Hyperbolic Transport]\label{r:cp}
This last example can be made even more obvious when abandoning the restriction to parabolic equations. Neglecting diffusion, the system (\ref{e:cp}) becomes a simple system of counter-propagating waves 
\[
u_t=u_x+\mu v,\ v_t=-v_x,\qquad  M_\lambda=\left(\begin{array}{cc} \lambda & \mu\\ 0 &-\lambda\end{array}\right).
\]
Fix $\mu\neq 0$. For $\lambda>0$ we have $E^\mathrm{u}_\lambda=\mathrm{Span}(1,0)^T$ and $E^\mathrm{s}_\lambda=\mathrm{Span}(\mu,-2\lambda)^T$, both of which define analytic families of subspaces in $\lambda\in\C$. However, since $E^\mathrm{s}_0=E^\mathrm{u}_0$, we see that $P^\mathrm{s}_\lambda$ is not analytic at $\lambda=0$. One readily finds that $P^\mathrm{s}_\lambda=\left(\begin{array}{cc} 0 &-\frac{\mu}{2\lambda} \\ 0&1\end{array}\right)$, which has a simple pole at $\lambda=0$. This is reflected in the fact that $2v(t,x)\to \int_{-\infty}^\infty u(y)\rmd y$ as $t\to\infty$, which is typically nonzero. On the other hand, for $\mu=0$, the stable subspace is simply $E^\mathrm{s}_\lambda=\mathrm{Span}(0,1)^T$, and $P^\mathrm{s}_\lambda=\left(\begin{array}{cc} 0 &0 \\ 0&1\end{array}\right)$ is analytic.
\end{Remark}

\section{Right-Sided Pointwise Growth}\label{s:2a}

In this section, we discuss a slightly different concept of pointwise stability. We think of a nonlinear growth process that has created a competing, more stable state, which now forms an effective boundary condition for the instability. We therefore study pointwise growth in problems on the half-line $x>0$ with some arbitrary boundary conditions at $x=0$, see Section \ref{s:3.1}.  We illustrate the relation to pointwise growth modes, continuing the previous examples, in Section \ref{s:3.2}, and briefly discuss relations to the Evans function in Section \ref{s:3.3}.  We will come back to the observations made here when discussing relations to nonlinear problems in Section \ref{s:dis}.

\subsection{Pointwise Growth on the Half-Line}\label{s:3.1}
We  consider the parabolic equation (\ref{e:01}) on the half line $x>0$, together with boundary conditions $B(u,\ldots,\partial_x^{2m-1}u)=0$, where $B:\C^{2mN}\to \C^{mN}$ is linear with full rank, with $mN$-dimensional kernel $E^\mathrm{bc}$. We assume that the boundary conditions give a well-posed system in the sense that $E^\mathrm{bc}\cap E^\mathrm{s}_\lambda=\{0\}$ for $\Re\lambda>0$, sufficiently large. As a consequence, we can define $P^\mathrm{bc}_\lambda$ as the projection along $E^\mathrm{bc}$ onto $E^\mathrm{s}_\lambda$. We also need the transported projection $P_\lambda^\mathrm{bc}(x)=\Phi_\lambda(x)P^\mathrm{bc}_\lambda\Phi_\lambda(-x)$. The Green's function associated with these boundary conditions is
\[
T_\lambda(x,y)=\begin{cases}
\Phi_\lambda(x-y)P_\lambda^\mathrm{bc}(y), & x>y\\
-\Phi_\lambda(x-y)(\mathrm{id}-P_\lambda^\mathrm{bc}(y)), & x<y.
\end{cases}
\]
In particular, $T_\lambda$ is analytic precisely when $P^\mathrm{bc}_\lambda$ is analytic. Clearly, singularities and pointwise growth may depend on the boundary conditions. One may wish to separate the influence of boundary conditions from properties of the medium. For this purpose, we can consider the subspace $E^\mathrm{s}_\lambda$ as a complex curve in the Grassmannian $\mathrm{Gr}_\C(2mN,mN)$ and discuss its singularities.
\begin{Definition}[BPGM and RPGM]\label{l:1s}
We refer to singularities of $P^\mathrm{bc}_\lambda$ as \emph{boundary right-sided pointwise growth modes} (BPGM) and to singularities of $E^\mathrm{s}_\lambda$ simply as right-sided pointwise growth modes (RPGM). The right-sided pointwise growth rate (RPGR) is defined as
the maximal real part of a right-sided pointwise growth mode, or,
\[
\inf \{\rho;\, E_\lambda^\mathrm{s}\mbox{ is analytic in }\Re\lambda>\rho\}.
\]
Considering $x<0$ and the unstable subspace $E^\mathrm{u}_\lambda$, one can define left-sided pointwise growth. 
\end{Definition}
We will now collect some properties of BPGMs and RPGMs.  First, we observe that right-sided pointwise growth modes are boundary right-sided pointwise growth modes. 
\begin{Lemma}[RPGM $\Rightarrow$ BPGM]\label{l:sing}
Singularities of $E^\mathrm{s}_\lambda$ are singularities of $P^\mathrm{bc}_\lambda$.
\end{Lemma}
\begin{Proof}
Suppose $P^\mathrm{bc}_\lambda$ is analytic. Then the range and kernel are analytic, in particular $E^\mathrm{s}_\lambda $ is analytic. 
\end{Proof}
There is a partial converse to Lemma \ref{l:sing} when we allow more general, dynamic, boundary conditions $u\in E^\mathrm{bc}_\lambda$.
\begin{Lemma}[BPGM vs RPGM]\label{l:sieq}
Suppose that $E^\mathrm{s}_\lambda$ is analytic in the open region $\mathcal{D}\subset \C$. Then there exists an analytic family of boundary conditions $E^\mathrm{bc}_\lambda$ so that the associated projection $P^\mathrm{bc}_\lambda$ is analytic in $\mathcal{D}$. 
\end{Lemma}
\begin{Proof}
We need to find a complementary subspace $E^\mathrm{bc}_\lambda$  to $E^\mathrm{s}_\lambda$. Such analytic complements always exist provided the domain is a Stein space; see for instance \cite[Thm 1]{shu} and references therein. Open subsets of $\C$ are Stein manifolds by the Behnke-Stein theorem; see for instance \cite{kra}.
\end{Proof}
The following lemma clarifies the relation between right-sided pointwise growth and pointwise growth. 
\begin{Lemma}[RPGM $\Rightarrow$ PGM]\label{l:frpgmpgm}
Singularities of $E^\mathrm{s}_\lambda$ are singularities of $P^\mathrm{s}_\lambda$. 
\end{Lemma}
\begin{Proof}
Analyticity of the projection $P^\mathrm{s}_\lambda$ implies analyticity of its range $E^\mathrm{s}_\lambda$.
\end{Proof}
The converse is true generically (see the discussion in Section \ref{s:alg}) but not true in general; see the example on counter-propagating waves, below. 

In order to determine analyticity of the subspace $E^\mathrm{s}_\lambda$, one can use local charts in the Grassmannian, for instance writing the subspace as a graph over a reference subspace, effectively embedding the Grassmannian locally into $\C^{mN\times mN}$. Alternatively, one can use the Pl\"ucker embedding into differential forms, working globally, albeit in a high-dimensional space.

\subsection{Examples --- continued}\label{s:3.2}

It is useful to recall our original motivation.  Suppose we are given a parabolic equation of the form (\ref{e:01}) with pointwise growth rate determined from a maximal pointwise growth mode.  We now restrict ourselves to the positive half-line, imposing boundary conditions at $x=0$ is such a way that the initial value problem is well-posed.  The question is whether boundary conditions can be selected so that the right-sided pointwise growth rate is strictly less than the pointwise growth rate.  That is -- can boundary conditions be selected so that faster pointwise rates of decay are observed for the problem on the half-line than for the problem on the whole real line?  The answer is given in the lemmas above.  To be precise, let $\lambda^*$ be the pointwise growth mode with maximal real part.  Then if $\lambda^*$ is also a RPGM then Lemma~\ref{l:sing} implies that $\lambda^*$ is also a BPGM and faster rates of decay are not possible by selecting appropriate boundary conditions.  On the other hand, if $\lambda^*$ is a PGM but not a RPGM then Lemma~\ref{l:sieq} guarantees that boundary conditions exist for which faster rates of decay are observed.  

We now return to the series of examples introduced in Section~\ref{sec:examples}.  To begin, we compute the RPGMs for the convection-diffusion and Cahn-Hilliard examples, showing the RPGMs always coincide with PGMs in these two examples.  Next we consider the counter-propagating waves example.  This is a particularly rich example that demonstrates that PGMs, RPGMs and BPGMs are not necessarily equivalent.  Having computed the RPGM, we will turn our attention to finding suitable boundary conditions such that faster pointwise rates of decay are observed on the half-line than for the same problem on the whole real line.

\paragraph{Convection-Diffusion.}
The stable subspace is given by $\mathrm{Span}(1,\mu_-(\lambda))^T$, which possesses a singularity at the pointwise growth mode $\lambda=-1/4$. 

\paragraph{Cahn-Hilliard.}
For $\mu>0$, the two-dimensional subspace possesses a singularity at $\lambda=\mu/2$. In fact, the subspace is spanned by $(1,\nu_j,\nu_j^2,\nu_j^3)^T$, where $\nu_j,j=1,2$ are the stable eigenvalues from (\ref{eq:stableCH}). At $\lambda=\mu/2$, $\nu_j=\pm\rmi\sqrt{\mu/2}$ are distinct, so that we can write the subspace as a graph from $(*,*,0,0)^T$ into $(0,0,*,*)^T$, represented by the square matrix
\[
L^\mathrm{s}_\lambda=\left(\begin{array}{cc}
-\nu_1\nu_2&\nu_1+\nu_2\\
-\nu_1\nu_2(\nu_1+\nu_2) & \nu_1^2+\nu_1\nu_2+\nu_2^2
\end{array}
\right),
\]
which is not analytic since $\nu_1+\nu_2$ is not analytic. One can similarly see that all other pointwise growth modes correspond to singularities of $E^\mathrm{s}_\lambda$ and therefore all PGMs are RPGMs (and therefore BPGMs).

\paragraph{Counter-Propagating Waves.}
We computed the stable subspace in (\ref{e:esu}). We can write the stable subspace globally as a graph over $(*,0,*,0)^T$ with values in $(0,*,0,*)^T$, which gives the matrix representation 
\[
L^\mathrm{s}_\lambda=\left(\begin{array}{cc}
\nu_u^-&
0\\
\frac{\mu(\nu_u^- -\nu_v^-)}{2\nu_u^-}&
\nu_v^-
\end{array}
\right),
\]
We clearly see singularities where the diagonals are singular, i.e. at $\lambda=-1/4$. When $\mu\neq 0$, we also see a singularity when $\nu_u^-=0$, $\nu_v^-\neq 0$. However, such a singularity does not occur along the principal branch of the square root, so that in this case, the pointwise growth mode at $\lambda=0$ is \emph{not} a right-sided pointwise growth mode. 


When considering the counter-propagating waves example with the drift directions switched in both the $u$ and $v$ component, we see that $L^\mathrm{s}_\lambda$ is singular at $\lambda=0$, when $\nu_u^-=0$, $\nu_v^-\neq 0$. We therefore need to analyze the subspace in a different coordinate system of the Grassmannian. Writing the stable subspace as a graph over $(0,0,*,*)^T$ into $(*,*,0,0)^T$, we find the matrix representation

\[
L^\mathrm{s}_\lambda=\frac{2\nu_u^-}{\mu(\nu_v^--\nu_u^-)}\left(\begin{array}{cc}
-\nu_v^- & 1 \\ 
-\nu_u^-\nu_v^- & \nu_u^- 
\end{array}
\right),
\]
which shows analyticity near $\nu_v^-\neq 0$, $\nu_u^-=0$. As was the case above, the pointwise growth mode at $\lambda=0$ is not an RPGM.  

We note that $E^\mathrm{s}_\lambda$ is not continuous in $\mu$ at $\lambda=0$. A somewhat lengthy calculation shows that the right-sided pointwise growth modes need not be continuous when adding bidirectional coupling,
\begin{align}
u_t&=u_{xx}+u_x+\gamma v\nonumber\\
v_t&=v_{xx}-v_x+u\label{e:cc}.
\end{align}
As we noticed, at $\gamma=0$, right-sided pointwise growth modes are located at $\lambda=-1/4$. For $\gamma\neq 0$, small, right-sided pointwise growth modes are ``created'' at $\lambda=0$, located at $\lambda=\pm\sqrt{\gamma}$. Rather than exhibiting the lengthy calculations that reveal the singularity, we refer to Remark \ref{r:rpgmnotc} in the following section, where this fact is shown using the fact that for $\gamma\neq0$, the growth mode is in some sense simple.

\begin{Remark}[Hyperbolic Transport ctd.] 
Again, all of the above can be observed in the simpler example of hyperbolic transport from Remark \ref{r:cp}. The stable subspace is given through $E^\mathrm{s}_\lambda=\mathrm{Span}(\mu,-2\lambda)^T$ is analytic at $\lambda = 0$, both for $\mu\neq 0$ and for $\mu=0$, when  $E^\mathrm{s}_\lambda=\mathrm{Span}(0,1)^T$. Again, notice that the stable subspace is not continuous in $\mu$ at $\lambda=0$. 
\end{Remark}

\paragraph{Counter-Propagating Waves --- BPGM versus RPGM.}
Consider example (\ref{e:cc}) with $\gamma=0$.  We will impose boundary conditions at $x=0$ and investigate pointwise stability of the zero state.  We know that the example (\ref{e:cc}), with $\gamma=0$,  has a pointwise growth mode at $\lambda=0$ and therefore the dynamics are pointwise marginally stable on the whole real line.  We also observed that $\lambda=0$ is not a RPGM.  Based upon this, we expect to observe the following dynamics
\begin{itemize}
\item marginal pointwise stability on $x\in\R$;
\item exponential decay on $x\in\R^+$ for suitable boundary conditions; Lemma \ref{l:sieq}.
\end{itemize}
Since the stable subspace can be written as a graph over $(*,0,*,0)^T$, Dirichlet boundary conditions, $E^\mathrm{bc}=(0,*,0,*)^T$ are always transverse to $E^\mathrm{s}_\lambda$ and therefore guarantee exponential decay. 
Phenomenologically, compactly supported initial conditions in $u$ are transported towards the boundary $x=0$ where the Dirichlet condition causes exponential decay. The $v$ equation is dominated by transport away from the boundary, where $v=0$ is ``fed'' into the system through the boundary condition, which again causes exponential decay.

\begin{figure}[ht]
   \includegraphics[width=0.49\textwidth]{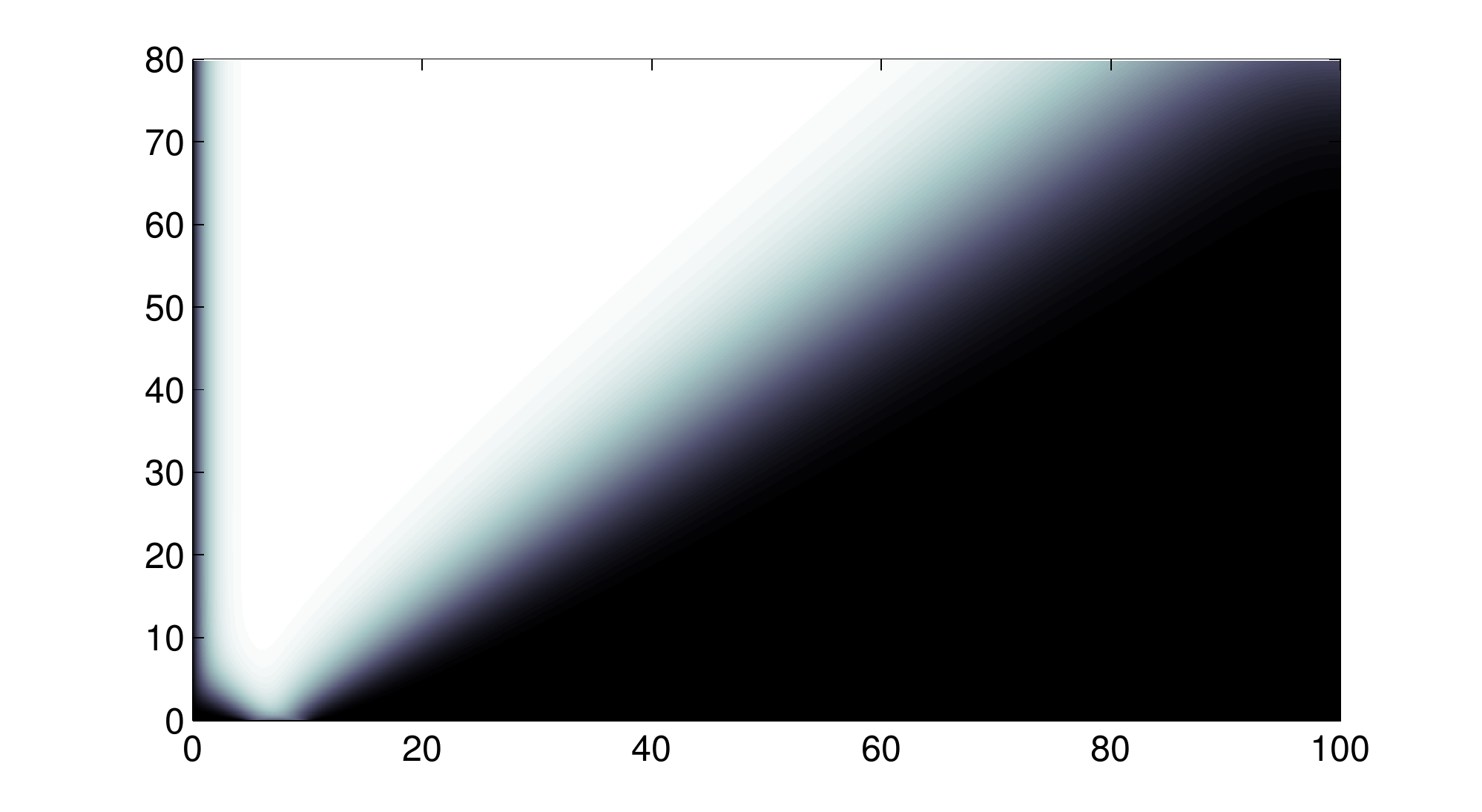}\hfill
    \includegraphics[width=0.49\textwidth]{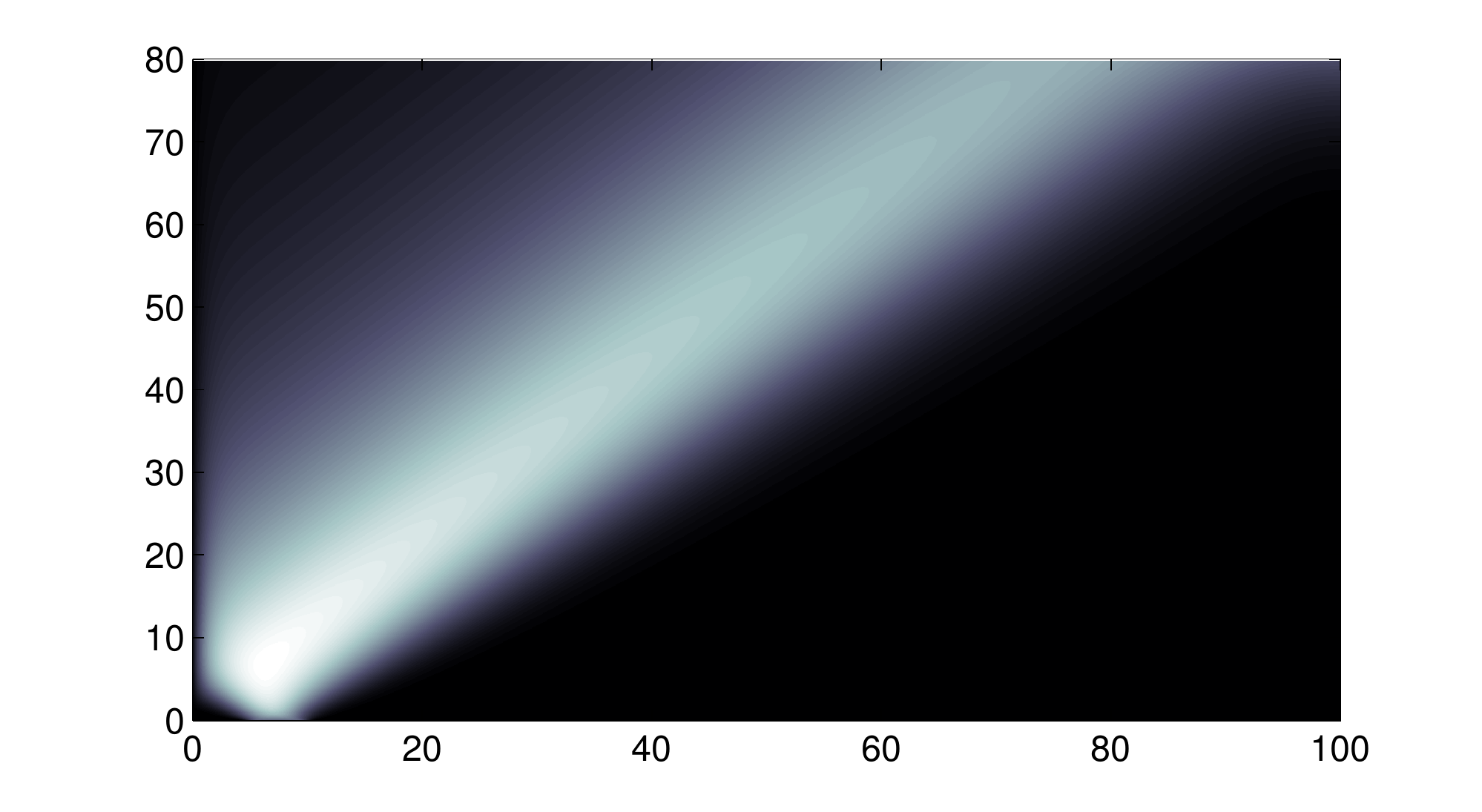}
\caption{Space-time plots of the $v$ component in (\ref{e:cc}).  On the left is a simulation of the case when $\kappa=0$ (Dirichlet boundary conditions).  On the right is a simulation with $\kappa=2$. When $\kappa=0$, the $v$ component remains nonzero.  For $\kappa=2$, pointwise exponential decay is observed with rate $e^{-t/4}$.     }
\label{fig:rightBC}
\end{figure}

Reversing the drift direction, that is, considering 
\begin{align}
u_t&=u_{xx}-u_x\nonumber\\
v_t&=v_{xx}+v_x+u\label{e:cc2},
\end{align}
we observe that the stable subspace at $\lambda=0$ is given by $(0,0,*,*)^T$, entirely contained in the $v$-equation. Consider boundary conditions at $x=0$ of the form $v=0$, $u=-\kappa v_x$.  It is straightforward to verify that these boundary conditions are transverse to $E^s_\lambda$ at $\lambda=0$.  Thus, $\lambda=0$ is not a BPGM.  The BPGMs can be determined explicitly from the singularities of 
\[
P^{bc}_\lambda=\left(\begin{array}{cccc}
\frac{-1+\sqrt{1+4\lambda}}{\kappa-1+\sqrt{1+4\lambda}}&
0 &
\kappa \frac{2\lambda}{\kappa-1+\sqrt{1+4\lambda}}&
\kappa \frac{-1+\sqrt{1+4\lambda}}{\kappa-1+\sqrt{1+4\lambda}} \\
-\frac{1}{2}\frac{(-1+\sqrt{1+4\lambda})^2}{\kappa-1+\sqrt{1+4\lambda}}&
0 &
-\kappa\frac{\lambda(-1+\sqrt{1+4\lambda})}{\kappa-1+\sqrt{1+4\lambda}}&
-\frac{\kappa}{2}\frac{(-1+\sqrt{1+4\lambda})^2}{\kappa-1+\sqrt{1+4\lambda}}\\
0 & 0 & 1 & 0\\
\frac{1}{\kappa-1+\sqrt{1+4\lambda}}& 0& 0\frac{2\lambda}{\kappa-1+\sqrt{1+4\lambda}}&
\frac{\kappa}{\kappa-1+\sqrt{1+4\lambda}}
\end{array}
\right).
\]
This projection has a singularity at $\lambda=\frac{(1-\kappa)^2-1}{4}$ for $\kappa\leq 1$ and we expect to observe pointwise exponential decay with this rate for any $\kappa>0$.  Phenomenologically, we solve the $v$-equation with homogeneous Dirichlet boundary conditions, observing transport to the boundary at $x=0$, with source $u$. Since such a system will decay exponentially in the absence of a source $u$ and we expect $v$ to relax to a constant, say $\sim\int u>0$, away from the boundary, introducing a boundary layer with positive slope. This in turn generates a \emph{negative} boundary source in the $u$-equation through $u=-v_x$, which then propagates to the right in the medium, generating a negative source in the $v$-equation. This will decrease the average of $v$ and therefore the slope of the boundary layer until eventually the system approaches zero locally uniformly exponentially. See Figure~\ref{fig:rightBC}.

We note that the immediate choice of transverse boundary conditions $v=v_x=0$ would make the system ill-posed. On the other hand, choosing separated boundary conditions, $a(u,u_x)=0, b(v,v_x)=0$, such as the Dirichlet boundary conditions above, would necessarily yield a non-zero intersection with the stable subspace that spans $(0,0,*,*)^T$.  This leads to a zero eigenvalue.  Still, we have demonstrated how appropriate choices of boundary conditions in the sense of Lemma \ref{l:sieq} can stabilize the system. More generally, our point here is that any instability would be generated by the boundary conditions in the sense that the pointwise growth rate depends non-trivially on the details of the boundary condition, unless it is simply $\lambda=-1/4$.

\begin{Remark}[Hyperbolic Transport ctd.] 
Without diffusion, negative coupling through the boundary can be achieved via $u=-v$ at $x=0$, which gives a boundary subspace $(1,-1)^T$ which is transverse to the stable subspace $(1,0)^T$ or $(2\lambda,1)^T$, respectively, in $\Re\lambda>-1/2$. One can also verify that the system is well-posed as the right-hand side generates a strongly continuous semi-group. 
\end{Remark}

\subsection{Pointwise Growth, Right-Sided Growth and the Evans Function}\label{s:3.3}

In the resolvent set, singularities of the Green's function correspond to eigenvalues and the order of the pole can be related to multiplicities. Zumbrun and Howard \cite{HZ} showed that this relation continues in a pointwise sense. That is, they showed that one can define projections on eigenvalues in a pointwise sense using the pointwise Green's function. Our situation here is much simpler since we assume translation invariance, so that the Green's function acts simply through convolution. On the other hand, our main interest here is in singularities of the Green's function caused by ``asymptotic'' stable and unstable eigenspaces which were excluded in \cite{HZ}.

Insisting on an Evans function approach, one could track $mN$-dimensional subspaces in $\C^{2mN}$ using differential forms. Differential forms solve an induced linear equation with equilibria corresponding to invariant subspaces. Singularities of the Green's functions are induced by two mechanisms: singularities of the stable or unstable subspaces, and intersections of stable and unstable subspaces. Right-sided pointwise growth modes track precisely the first type of singularities. Pointwise growth modes combine both types of singularities and singularities correspond to either branch points or roots of the Evans function. In an Evans function approach, one usually regularizes singularities of $E^\mathrm{s}_\lambda$ by going to appropriate Riemann surfaces and tracks intersections of $E^\mathrm{s}_\lambda$ and $E^\mathrm{u}_\lambda$ (or $E^\mathrm{bc}_\lambda$) by forming a wedge product of the differential forms associated with the two subspaces.

\section{Algebraic Criteria --- Pinched Double Roots}\label{s:alg}

In this section, we review the more common approach to pointwise stability based on pinched double roots of the dispersion relation. We compare pinched double roots with pointwise growth modes and we illustrate differences in examples.

\subsection{Pinched Double Roots and Algebraic Pointwise Growth}

The more common approach to stability and instability problems uses the Fourier transform to reduce the stability problem to a parameterized family of matrix eigenvalue problems, $(\lambda - A(\rmi k))u=0$, with roots precisely where
\[
d(\lambda,\rmi k):=\mathrm{det}\,( A(\rmi k) -\lambda\mathrm{id})=0.
\]
We call $d:\C^2\to\C$ the dispersion relation. There are $N$ roots $\lambda_j(\nu)$ of $d(\lambda,\nu)$ for fixed $\nu$, whereas there are $2mN$ roots $\nu_\ell(\lambda)$ for fixed $\lambda$. Using ellipticity of $A$, we find that for $\Re\lambda\to\infty$, there are precisely $mN$ roots $\nu_j$ with $\Re\nu_j(\lambda)\to +\infty$ and $mN$ roots $\nu_j$ with $\Re\nu_j\to -\infty$. Note that the roots $\nu_j$ are in general not analytic in $\lambda$. Non-analyticity can occur only when at least two of the roots coincide. This occurs in the case of multiple zeros of the dispersion relation. 

\begin{Definition}[Double Roots]\label{d:dr}
We say $(\lambda_*,\nu_*)$ is a double root of the dispersion relation if
\begin{equation}\label{e:dr}
d(\lambda_*,\nu_*)=0,\qquad \partial_\nu d(\lambda_*,\nu_*)=0.
\end{equation}
We say that $(\lambda_*,\nu_*)$ is a simple double root when $\partial_\lambda d(\lambda_*,\nu_*)\neq 0$ and $\partial_{\nu\nu}(\lambda_*,\nu_*)\neq 0$.
\end{Definition}
One readily verifies that simple double roots are simple as solutions to the complex system (\ref{e:dr}). 

\begin{Definition}[Pinching]\label{d:pi}
We say that a double root $(\lambda_*,\nu_*)$ is \emph{pinched} if there exists a continuous curve $\lambda(\tau)$, $\tau\in\R^+$ with $\Re\lambda(\tau)$ strictly increasing, $\lambda(0)=\lambda_*$, $\Re\lambda(\tau)\to\infty$ for $\tau\to\infty$,  and continuous curves of roots $\nu_\pm(\lambda(\tau))$ to $d(\lambda(\tau),\nu_\pm(\lambda(\tau)))=0$ so that
\[
\nu_\pm(\lambda_*)=\nu_*,\qquad \lim_{\tau\to\infty}\Re\nu_\pm(\lambda(\tau))=\pm\infty.
\]
\end{Definition}
In analogy to algebraic and geometric multiplicities of eigenvalues, one can think of pinched double roots as \emph{algebraic} pointwise growth modes, as opposed to \emph{geometric} pointwise growth modes that characterize singularities of the Green's function. We then refer to the largest real part of a pinched double root as the \emph{algebraic pointwise growth rate}. 

\subsection{Pointwise Growth versus Pinched Double Roots}
The following lemmas and remarks relate pinched double roots and pointwise growth modes. 
\begin{Lemma}[PGM $\Rightarrow$ PDR]\label{l:p-p}
Let $\lambda_*$ be a pointwise growth mode. Then there exists $\nu_*$ so that $(\lambda_*,\nu_*)$ is a pinched double root. 
\end{Lemma}
\begin{Proof}
Suppose there are no pinched double roots with $\lambda=\lambda_*$. Then the subspaces $E^\mathrm{s}_\lambda$ and $E^\mathrm{u}_\lambda$ correspond to different spectral sets of $M_\lambda$ and can therefore be continued in an analytic fashion as complementary eigenspaces, contradicting the assumption of a pointwise growth mode with $\lambda=\lambda_*$. 
\end{Proof}
Together with Lemma \ref{l:frpgmpgm}, this gives RPGM $\Rightarrow$ PGM $\Rightarrow$ PDR. 
\begin{Lemma}[Simple PDR $\Rightarrow$ RPGM]\label{l:p-p1}If $(\lambda_*,\nu_*)$ is a simple pinched double root, then $E^\mathrm{s}_\lambda$ is not analytic at $\lambda_*$.
\end{Lemma}
\begin{Proof}
In a simple pinched double root, we have expansions $\nu_\pm\sim \pm\sqrt{\lambda}$, where we assumed without loss of generality that $\lambda_*=\nu_*=0$. Because $\partial_\lambda d\neq 0$, the eigenvectors to the eigenvalues $\nu_\pm$ become co-linear at $\lambda=0$, since otherwise $M_0$ would have a two-dimensional null-space and $d(\lambda,0)=\rmO(\lambda^2)$. If the eigenspace $E^\mathrm{s}_\lambda$ were analytic, we could trivialize it locally by an analytic change of coordinates and consider the eigenvalue problem within this one-dimensional eigenspace, which clearly guarantees analyticity of the eigenvalue, thus contradicting the presence of a simple pinched double root. As a consequence, the stable subspace is not analytic.
\end{Proof}
Together with Lemma \ref{l:frpgmpgm}, this also gives ``simple PDR'' $\Rightarrow$ PGM.

%
\begin{Remark}[PDR $\not\Rightarrow$ PGM]\label{r:not}
We will see in the example of counter-propagating waves, below that the assumption of a simple pinched double root is indeed necessary. In particular, there are pinched double roots that do not correspond to pointwise growth modes. In other words, pinched double roots may overestimate pointwise growth rates, albeit only in non-generic cases, when pinched double roots are not simple. 
\end{Remark}

\begin{Remark}[RPGMs need not be continuous]\label{r:rpgmnotc}
We can use Lemma \ref{l:p-p1} to see that right-sided pointwise growth modes are not continuous. In (\ref{e:cc}), there is a double pinched double root at $\lambda=\nu=0$ for $\mu=0$, which does not give rise to a singularity of the stable subspace. Upon perturbing to $\mu\neq 0$, we find two simple double roots at $\lambda=\pm\sqrt{\mu}$, $\nu=\rmO(\mu)$, and therefore a one-sided pointwise growth mode. 
\end{Remark}

\subsection{Examples --- continued}
We compute pinched double roots for the examples in Section~\ref{sec:examples} and contrast them with the singularities found in the pointwise Green's function.  We emphasize the example of counter-propagating waves, which possesses a pinched double root without having a pointwise growth mode at $\lambda=0$. 

\paragraph{Convection-Diffusion.} The dispersion relation and its derivative with respect to \(\nu\) are
\begin{align} \label{dispconvdiff}
d(\lambda,\nu) &= \nu^2+\nu-\lambda \nonumber\\
\partial_{\nu}d(\lambda,\nu) &= 2\nu + 1.
\end{align}
Setting both equations equal to zero we find there is only one double root at $ \lambda=-1/4, \,\nu=-1/2 $. Writing the roots of the dispersion relation as
\[ \nu_\pm(\lambda)=-\frac{1}{2}\pm\frac{1}{2}\sqrt{1-4\lambda},\]
we see that the double root occurs when the terms inside the square root vanish.  Taking $\lambda\to\infty$ we find that this double root is pinched.  Note that the pointwise Green's function derived in Section~\ref{sec:examples} has a singularity at precisely this point.  Thus, in this example pinched double roots and algebraic pointwise growth modes are equivalent, according to Lemmas \ref{l:p-p} and \ref{l:p-p1}.

\paragraph{The Cahn-Hilliard Linearization.} The dispersion relation and its derivative with respect to \(\nu\) are
\begin{align} \label{dispcahnhill}
d(\lambda,\nu) &= \nu^4+\mu\nu^2+\lambda \nonumber\\
\partial_{\nu}d(\lambda,\nu) &= 4\nu^3+2\mu\nu.
\end{align}
The double roots occur for $(\lambda,\nu)=(0,0)$ and $(\lambda,\nu)=(\mu^2/4,\pm \rmi\sqrt{\mu/2})$.  From \(d(\lambda,\nu) = 0\) we  find four roots of the dispersion relation,
\[
\pm\sqrt{\frac{-\mu\pm\sqrt{\mu^2-4\lambda}}{2}}.
\] 
When $\mu>0$, then the two stable roots are
\[
\sqrt{\frac{-\mu-\sqrt{\mu^2-4\lambda}}{2}} \text{  and } -\sqrt{\frac{-\mu+\sqrt{\mu^2-4\lambda}}{2}}.
\] 
Recall that we have taken the principle value of the square root to lie in the upper half of the complex plane.  Owing to this, when $\mu>0$ the double root at $(\lambda,\nu)=(\mu^2/4,\rmi\sqrt{\mu/2})$ involves the roots $\sqrt{\frac{-\mu\pm\sqrt{\mu^2-4\lambda}}{2}}$ and is pinched.  Therefore, when $\mu>0$ there exist algebraic and pointwise growth modes in the right half plane.  

When $\mu<0$, the two roots involved in the double root at $(\lambda,\nu)=(\mu^2/4,-\rmi\sqrt{\mu/2})$ are the two stable roots above and therefore not pinched.  

Again, pinched double roots and pointwise growth modes (\ref{eq:ptGCH}) coincide for $\mu\neq 0$, in accordance with Lemmas \ref{l:p-p} and \ref{l:p-p1}. 

Note also that for $\mu=0$, there is a multiple pinched double root at $\lambda=\nu=0$. More precisely, $\partial_{\nu\nu}d$=0 at this double root. There also is a pointwise growth mode at $\lambda=0$, although Lemma \ref{l:p-p1} does not guarantee the existence. 

\paragraph{Counter-Propagating Waves.} The dispersion relation and its derivative with respect to \(\nu\) are
\begin{align} \label{dispcounterl}
d(\lambda,\nu) &= (\nu^2-\lambda)^2-\nu^2\nonumber\\
\partial_{\nu}d(\lambda,\nu) &= 2\nu(2(\nu^2-\lambda)-1)
\end{align}
The double roots are at $ \lambda=0,\, \nu=0 ;\, \lambda=-1/4,\, \nu=-1/2$ and $\lambda=-1/4,\, \nu=1/2  $. They all pinch for any value of $\mu$. However, from the projection $P_\lambda^\mathrm{s}$, we notice that $\lambda=0$ is no longer a pointwise growth mode for $\mu=0$.  This example shows that pinched double roots may not give rise to pointwise growth modes, as announced in Remark \ref{r:not}.

\begin{Remark}\label{r:cont} 
We observe in this example that pinched double roots ``resolve'' the discontinuity observed in Remark \ref{r:usc}: pinched double roots are continuously depending on $\mu$ in this example, while a pointwise growth mode disappears at $\mu=0$ and the pointwise growth rate jumps.
\end{Remark}

\section{Properties of Pointwise Growth Modes}\label{s:prop}
In this section, we give rough $\lambda$-bounds, existence results and some counting results for pointwise double roots and pointwise growth modes, Section \ref{s:5.1}, and study continuity properties in Section \ref{s:5.2}.  

\subsection{Existence, Counts, and Bounds}\label{s:5.1}

Recall that the operator $A(\partial_x) - \lambda$ is invertible on, say, $L^2(\R,\C^N)$ if and only if $A(\rmi k)-\lambda$ is invertible for all $k\in\R$. Then $\sigma (A)$ is given by the zero set of $d(\cdot,\rmi k)$, $k\in\R$. Similarly, we found spectra in exponentially weighted spaces $\sigma_\eta (A)$ from roots of $d(\cdot,\rmi k+\eta)$. 

\begin{Lemma}\label{l:ess}
Pinched double roots are bounded to the right by the essential spectrum. More precisely, for any algebraic pointwise growth mode $\lambda_*,\nu_*$, the pinching path $\lambda(s)$ necessarily intersects the essential spectrum $\sigma_\eta (A)$ for any $\eta\in\R$. 
\end{Lemma}
\begin{Proof}
From an algebraic pointwise growth mode, we can follow $\nu_\pm(\lambda(s))$ as $\lambda(s)\to +\infty$. Since $\nu_+(\lambda(0))=\nu_-(\lambda(0))$ and $\Re\nu_\pm\to\pm\infty$, there is $s_*\geq 0$ so that either $\Re\nu_+=0$ or $\Re\nu_-=0$. Since $d(\lambda(s_*),\nu_\pm(\lambda(s_*)))=0$, $\lambda(s_*)\in\sigma_\mathrm{ess}$. 
\end{Proof}
Of course, bounds on pinched double roots imply bounds on pointwise growth modes and right-sided pointwise growth modes by Lemmas \ref{l:frpgmpgm}  and \ref{l:p-p}.

Exploiting ellipticity, we can show that $\sigma_\mathrm{ess}$ (and therefore all pinched double roots) is contained in a sector $|\Im\lambda|\leq C(M-\Re\lambda)$. 

The simple example $u_t=u_x$ shows that not all well-posed PDEs possess algebraic pointwise growth modes. On the other hand, the class of parabolic equations that we have focused on do.
\begin{Lemma}\label{l:ex}
The parabolic PDE (\ref{e:0}) possesses at least one finite (right-sided) pointwise growth mode. 
\end{Lemma}
\begin{Proof}
We show that the projection $P^\mathrm{s}_\lambda$ cannot be analytic on $\C$ and conclude from the construction that the range cannot be analytic either. We therefore compute the leading order expansion for $|\lambda|$ large and show that the resulting function cannot be an analytic function of $\lambda$. More precisely, we find that the analytic continuation of $P^\mathrm{s}_\lambda$ along the circle $\lambda=R\rme^{\rmi\varphi}$, $R$ large, fixed,  does not result in a univalent function, indicating at least one branch point singularity inside the circle. 

For this, we expand the dispersion relation 
\[
d(\lambda,\nu)=\mathrm{det}\,( A(\nu) -\lambda \mathrm{id})
\]
in $|\lambda|$, setting $\lambda=R\rme^{\rmi\varphi}$ and $\nu=R^{1/2m}\hat{\nu}$, to find
\[
d(\lambda,\nu)=R^N\mathrm{det}\,\left(A_{2m}\hat{\nu}^{2m} (1+\rmO(R^{-1/2m})) - \rme^{\rmi\varphi} \mathrm{id}\right).
\]

Denote by $\rho_j$, $j=1,\ldots,N$ the eigenvalues of $A_{2m}$.  The roots  $\nu_j$ are therefore given as 
\[
\nu_{j,\ell}=R^{1/2m} \rho_j^{-1/2m} \rme^{2\pi\rmi\ell/2m}\rme^{\rmi\varphi/2m}(1+\rmO(R^{-\gamma})),\quad 1\leq j\leq N,\ 0\leq \ell < 2m.
\]
Here, we use the branch of the $2m$'th root fixing the positive real line. Ellipticity guarantees that $(-1)^m\Re\rho_j>0$, which then readily implies that for $\varphi=0$ there are precisely $mN$ roots with $\Re\nu>0$ and $mN$ roots with  $\Re\nu<0$, thus giving rise to projections $P^\mathrm{s/u}(\lambda)$ on $\lambda>0$. The set of roots corresponding to $P^\mathrm{s}_\lambda$ can be continued in $\varphi$ for all $\varphi\in\R$, so that the projections possess an analytic extension via Dunford's integral. Inspecting the formula for the $\nu_{j,\ell}$, we notice however that for $\varphi=2\pi m$, the roots $\nu$ are multiplied by $-1$, approximately, and therefore $P^\mathrm{s}_\lambda$, given by the analytic continuation from $R$ to $R\rme^{2\pi \rmi m}$, equals $P^\mathrm{u}_\lambda$ at $\lambda=R$. As a consequence, $P^\mathrm{s}_\lambda$ is not analytic in $\lambda$, in fact not even well-defined on $\C$. This shows existence of a singularity of $P^\mathrm{s}_\lambda$ and the existence of a pointwise growth mode as claimed. Clearly, the range of $P^\mathrm{s}_\lambda$ gives the analytic extension of the stable subspace $E^\mathrm{s}_\lambda$, which therefore cannot be analytic, either. 
\end{Proof}

Counting algebraic growth modes is difficult because of the pinching condition. The following example shows that the number of algebraic growth modes can actually jump. 

\begin{Example}
Consider again $u_t=-u_{xxxx}-\mu u_{xx}$ with dispersion relation $d(\lambda,\nu)=\lambda+\nu^4+\mu\nu^2$. One readily finds double roots at 
\[
(\lambda,\nu)=\left\{\begin{array}{l}
(0,0)\\
(\mu^2/4,\sqrt{-\mu/2})\\
(\mu^2/4,-\sqrt{-\mu/2}).
\end{array}
\right.
\]
For $\mu>0$, one can readily see that the purely imaginary pair of roots $\nu_\pm=\pm\rmi \sqrt{|\mu|/2}$ split off the imaginary axis and converge to $\pm\infty$ as $\lambda>\mu^2/4$ is increased to infinity, so that those two roots are pinched. On the other hand, the two roots $\nu\sim 0$ that create the double root at $\lambda=0$ split into two imaginary roots as $\lambda>0$ increases and eventually meet, separately, other roots when $\lambda$ passes $\mu^2/4$, and it is impossible to continue the specific root beyond this point. However, choosing any path from $\lambda=0$ to $\lambda=+\infty$ off the real axis, one sees that the roots initially split with opposite real parts. Since the path does not cross the (real) essential spectrum, the roots actually pinch along any such path. In summary, there are two pinched double roots, for all $\mu>0$, and the third double root pinches along any non-real path. 

For $\mu<0$, one readily sees that $\lambda=0$ pinches. The other two double roots, however, cannot pinch since the essential spectrum is to the left of the double root. Pinching paths for these double roots would need to pass around the origin. 

In summary, there are two pinched double roots for $\mu>0$ and only one pinched double root for $\mu<0$. 
\end{Example}

One can however count double roots in general. General double roots are solutions to the system of polynomial equations $d=0,\partial_\nu d=0$, which in turn can be counted using B\'ezout's theorem or resultants. Exact counts need to take multiple roots into account, where multiplicity can be defined via an algebraic intersection number  (see \cite[\S 5.3]{fulton}) or topologically via Brower's degree.

\begin{Lemma}\label{l:res}
Consider a system of $N$ strictly parabolic equation of order $2m$ (\ref{e:01}) and assume that the eigenvalues of $A_{2m}$ are all algebraically simple. Then the dispersion relation $\mathrm{det}\,(A(\nu)-\lambda)$ possesses precisely $2mN^2-N$ double roots, counted with multiplicity. 
\end{Lemma}
\begin{Proof}
The assumption on the eigenvalues of $A(\nu)$ guarantees that double roots are uniformly bounded for bounded matrices $A_j$, $1\leq j\leq 2m-1$. One can therefore perform a homotopy to a simple, homogeneous equation with $d(\lambda,\nu)=\prod_{j=1}^N (\lambda-\rho_j\nu^{2m}-a_j)$, for which double roots can be computed explicitly as follows. First, double roots of multiplicity $2m-1$ arise when $\nu=0,\lambda=a_j$. Second, double roots of multiplicity 2 arise for $\lambda$ values when two factors are equal, which occurs when $(\rho_j-\rho_{j'})\nu^{2m}=a_{j'}-a_{j}$, which yields a total of $N(N-1)2m$ roots. Together with the double roots at $\lambda=a_j$, we find the desired total of $2mN^2-N$. 
\end{Proof}
We remark that the result also follows from a direct computation of the resultant of $d(\lambda,\nu)$ and $\partial_\nu(\lambda,\nu)$ with respect to the variable $\nu$, or from B\'ezout's theorem \cite{fulton}. In fact, substituting $\lambda=\gamma^{2m}$ in the dispersion relation, one finds that there are no intersections at infinity, which assures that the number of roots (in $\gamma$) is given by the product of the degrees of $d$ and $\partial_\nu d$, $2mN\cdot (2mN-1)$. Dividing by $2m$ then gives the number of roots $(\lambda,\nu)$. 

\begin{Example}

\begin{enumerate}
\item Consider again the fourth-order evolution $u_t=-u_{xxxx}+\mu u_{xx}$ with dispersion relation $\lambda+\nu^4-\mu\nu^2=0$. The derivative $\partial_\nu d =0$ yields precisely three double roots $\nu$, counted with multiplicity, which corresponds to $2mN^2-N$ at $N=1,m=2$.
\item Multiple eigenvalues of $A_{2m}$ can produce continua of double roots, as the example of scalar diffusion in $\R^N$, $d(\lambda,\nu)=(\lambda-\nu^2)^N$ with double roots $\lambda=\nu^2\in\C$, arbitrary, shows. 
\item For $d(\lambda,\nu)=(\lambda-\nu^2+2a\nu+b)(\lambda-\nu^2)$, corresponding to a system of convection-diffusion equations with scalar diffusion, one finds double roots at $\nu=a,\nu=0$, and a double double root at $\nu=-b/2a$. In particular, there are 4 double roots for $a\neq 0$, 2 double roots for $a=0,b\neq 0$, and infinitely many double roots for $a=b=0$. 
\end{enumerate}
\end{Example}

\subsection{Continuity and Robustness}\label{s:5.2}

The following lemma establishes that the sudden increase in the pointwise growth rate upon arbitrarily small perturbations as exemplified in Remark \ref{r:usc} is the only type of discontinuity that can occur.
\begin{Lemma}\label{l:jump}
Pointwise growth rates are lower semi-continuous. 
\end{Lemma}
\begin{Proof}
We need to show that pointwise growth modes are robust. More specifically, we show that they can not disappear under arbitrarily small perturbations. Pointwise growth modes correspond to singularities of the projection $P^\mathrm{s}_\lambda$. These possess Puisseux-expansions near singularities. In particular, the winding number of at least one coefficient of $P^\mathrm{s}_\lambda$ is not a positive integer, also for small perturbations. As a consequence, $P^\mathrm{s}_\lambda$ cannot be analytic for nearby systems in a neighborhood of a singularity. 
\end{Proof}

Similar difficulties occur when considering right-sided pointwise growth rates. We mentioned in the example of counter-propagating waves in Section \ref{s:3.2} that right-sided pointwise growth rates need not be continuous in system parameters; see also Remark  \ref{r:rpgmnotc}, below. The following lemma establishes lower semi-continuity. 
\begin{Lemma}\label{l:jump2}
Right-sided pointwise growth rates are lower semi-continuous. 
\end{Lemma}
\begin{Proof}
Similar to the proof of Lemma \ref{l:jump}, we suppose that $E^\mathrm{s}_\lambda$ possesses a singularity at $\lambda=\lambda_*$. By compactness of the Grassmannian, there exists an accumulation point $E^\mathrm{s}_{\lambda_*}$. Near $\lambda_*$, the stable subspace possesses a Puisseux expansion and the winding number of at least one coefficient is not a positive integer, a fact that persists upon perturbing. 
\end{Proof}

Algebraic pointwise growth rates are easier to control, as the following result shows. 

\begin{Lemma}\label{l:robust}
Assume that there there are finitely many double roots. Then algebraic pointwise growth rates are robust. More precisely, consider a parameterized family of operators $A(\partial_x;\mu)$ such that the coefficients of the differential operator depend continuously on $\mu$. Then the algebraic pointwise growth rate is continuous in $\mu$. 
\end{Lemma}
\begin{Proof}
Assume that $(\lambda_*,\nu_*)$ is a pinched double root at $\mu=0$. Without loss of generality, assume $\lambda_*=\nu_*=0$. We claim that there is a pinched double root nearby for $\mu$ sufficiently small. For all $\mu$, there are finitely many double roots in a small neighborhood of the origin that all converge to the origin as $\mu \to 0$. At $\mu=0$, we have finitely many roots $\nu_j$, $j=1,\dotsc,J$, of the dispersion relation $d(\lambda,\nu)=0$ that converge to the origin as $\lambda\to 0$. Also, there is $M$ so that the roots $\nu_j$ are analytic functions $\nu_j(\gamma)$, when $\gamma=\lambda^{1/M}$. Here, we again use the standard cut of the square root, so that $\gamma>0$ when $\lambda>0$. 

Since we assume that the origin is a pinched double root, we have that, say, $\Re\nu_j(\gamma)\to+\infty$, $j=1,\ldots,J_+$ and $\Re\nu_j(\gamma)\to -\infty$, $j=J_++1,\ldots,J$ for $\gamma>0$ and $\gamma\nearrow\infty$.

Define 
\[
d_+(\gamma,\nu)=\prod_{j=1}^{J_+}(\nu-\nu_j(\gamma)),\qquad 
d_-(\gamma,\nu)=\prod_{j=J_++1}^{J}(\nu-\nu_j(\gamma)).
\]
Clearly, double roots correspond to solutions of the analytic system of equations in two variables $d_\pm(\gamma,\nu)=0$. Note that $\nu=\gamma=0$ is an isolated solution to this system, since otherwise an analytic function $\nu_j(\gamma)-\nu_{j'}(\gamma)$ would vanish identically for some $j \leq J_+< j'$, contradicting the fact that $\lambda_*$ was a pointwise growth rate. As a consequence,  the Brower degree is positive. 

Next, consider $d_\mu(\gamma^M,\nu)=0$ with $\mu$ sufficiently small. For $\gamma>0$, not too small, there are $J_+$ roots $\nu_j(\gamma;\mu)$ which converge to $+\infty$ as $\gamma\to+\infty$. Define 
\[
d_+^\mu(\gamma,\nu)=\prod_{j=1}^{J_+}(\nu-\nu_j(\gamma;\mu)),\qquad 
d_-^\mu(\gamma,\nu)=\prod_{j=J_++1}^{J}(\nu-\nu_j(\gamma;\mu)).
\]
We now argue by contradiction. We assume that there are no pinched double roots in a neighborhood of $\lambda=\nu=0$ for values of $\mu$ arbitrarily small. Under this assumption, we claim that $d_\pm^\mu$ are analytic. To see this, first notice that for $\gamma$ small, the distance between roots is bounded from below,
\[
\inf_\gamma \min_{j\leq J_+,j'>J_+} |\nu_j(\gamma)-\nu_{j'}(\gamma)|>\delta>0.
\]
We now choose a  family of curves $\Gamma(\gamma)$ as the boundary of a union of small balls around the roots of interest,
\[
\Gamma(\gamma)=\bigcup_{j=1}^{J_+} B_{\delta_2}(\nu_j(\gamma)).
\]
We then define the analytic functions
\[
M_\ell(\gamma)=\frac{1}{2\pi\rmi}\int_{\Gamma(\gamma)}\nu^\ell \frac{\partial_\nu d_\mu(\gamma^M,\nu)}{d_\mu(\gamma^M,\nu)}\rmd\nu.
\]
From Cauchy's theorem, we find that $M_\ell$ are the symmetric power sum polynomials in the eigenvalues,
\[
M_\ell(\gamma)=\sum_{j=1}^{J_+}\nu_j^\ell(\gamma).
\]
Using Newton's identities, we can express all elementary symmetric polynomials in terms of power sums. In other words, the $M_\ell$ generate the ring of symmetric polynomials.   The coefficients of $d_+^\mu(\gamma,\nu)$, viewed as a polynomial in $\nu$, are symmetric polynomials and can therefore be expressed in terms of the $M_\ell$, which shows that $d_+^\mu$ is analytic in $\gamma$ and $\nu$. Completely analogously, we find that $d_-^\mu(\gamma,\nu)$ is analytic. By assumption, we know that the complex system $d_\pm^\mu(\gamma,\nu)=0$ does not possess solutions for $\gamma$ close to zero.

We will now produce a contradiction by showing that $d_\pm^\mu$ and $d_\pm$ are homotopy equivalent on the boundary of a small neighborhood $B(0)$ of $\gamma=\nu=0$, that is, there is a homotopy between the two equations that does not possess roots on the boundary. We specifically consider the boundary of $|\nu|\leq \delta_\nu$, $|\gamma|\leq \delta_\gamma$, with $\delta_\nu,\delta_\gamma$ small. We decompose the boundary into two parts, where $I=\{|\nu|=\delta_\nu,|\gamma|\leq\delta_\gamma\}$ and $II=\{|\gamma|=\delta_\gamma,|\nu|\leq\delta_\nu\}$, respectively. 

In region $I$, we fix $\delta_\nu$ sufficiently small and notice that, as $\mu,\delta_\gamma \to 0$, $d^\mu_\pm$ and $d_\pm$ converge to $\nu^J$ in region $I$. As a consequence, for $\delta_\gamma$ sufficiently small, $d_\pm$ and $d_\pm^\mu$ are homotopy equivalent via a straight homotopy on region $I$. 

With $\delta_\gamma$ and $\delta_\nu$ fixed as above, we now consider region $II$. The groups of roots $\{\nu_j; \ 1\leq j\leq J_+\}$ and $\{\nu_j; \ J_+< j\leq J\}$  are well defined and separated by a finite distance for $\mu=0$. The construction of $d_\pm^\mu$ from above is therefore continuous in $\mu$, so that $d_\pm^\mu$ is continuous in $\mu$ on $II$. The same reasoning applies to $d_-$ and $d_-^\mu$.

Taken together, we conclude that $d_\pm$ and $d_\pm^\mu$ possess the same degree on $\partial B(0)$, a contradiction to the fact that the degree of $d_\pm^\mu$ vanishes and the degree of $d_\pm$ does not. 

This concludes the proof of continuity of algebraic pointwise growth modes. \end{Proof}

\begin{Remark}\label{r:p=p}
Using Thom transversality one can see that generically all double roots are simple. By Lemma \ref{l:p-p1}, simple pinched double roots are pointwise growth modes, which implies algebraic pointwise growth rates generically equal pointwise growth rates and pointwise growth rates generically are continuous. 
\end{Remark}

\section{Spreading Speeds}\label{s:sp}

In this section we exploit pointwise stability concepts to characterize spatial spreading of instabilities. Some of the following definitions and results are contained in \cite{ssmorse,tina} but will be repeated here to make the discussion more accessible. 

We are interested in unstable states, $\Re\sigma_\mathrm{ess}\cap \{\Re\lambda>0\}\neq\emptyset$. Instability of the spectrum implies that localized perturbations will grow exponentially in the $L^2$-norm. However, they may decay in a localized window of observation. Slightly generalizing from the previous discussion, we now allow this window of observation to move with speed $s$.  That is, we consider (\ref{e:0}) in a comoving frame of reference $\xi=x-st$ and study pointwise growth depending on $s$. In the following, we choose to rely on pinched double roots as criteria for pointwise growth. As we saw, pinched double roots may overestimate pointwise growth rates. On the other hand, pinched double roots are technically easier to work with, giving in particular continuity of growth rates. We are interested in the set of speeds $s$ for which there are pinched double roots in $\Re\lambda>0$ of the dispersion relation in a comoving frame, 
\begin{equation}\label{e:co}
d_s(\lambda,\nu):=d(\lambda-s \nu,\nu)=\mathrm{det}\,(A(\nu)+ s\nu-\lambda).
\end{equation}

\begin{Definition}[Spreading Speeds]\label{d:ss}
We say that $s_+$ is the spreading speed (to the right) of (\ref{e:0}) if
\[
s_+=\sup\{s; \ d_s \mbox{ possesses a pinched double root in } \Re\lambda\geq 0\}.
\]
\end{Definition}
Note that, by the previous discussion, the system will typically (that is, whenever the algebraic pointwise growth modes are pointwise growth modes) be pointwise unstable in frames with speed less than but arbitrarily close to $s_+$. In this context, it will be helpful to think of group velocities in a generalized fashion. 
\begin{Definition}[Group Velocities]\label{d:cg}
Let $d(\lambda_*,\nu_*)=0$ and $\partial_\lambda d(\lambda_*,\nu_*)\neq 0$. Then we define the group velocity as
\[
s_\mathrm{g}:=-\partial_\nu d/\partial_\lambda d_{|(\lambda_*,\nu_*)}.
\]
\end{Definition}
One readily verifies that $s_\mathrm{g}^\mathrm{co}=s_\mathrm{g}-s$ in a comoving frame $\xi=x-st$. Moreover, $s_\mathrm{g}=0$ implies $\partial_\nu d=0$ and hence the presence of a double root. Therefore, $\lambda_*$ is a double root in a suitable comoving frame when $s_\mathrm{g}$ is real.

\begin{Lemma}[Spreading Speeds \& Group Velocities]\label{l:max}
Let $\lambda_*$ be an element in the essential spectrum with extremal real part, that is, 
\begin{itemize}
\item $\lambda_*$ is simple, $d(\lambda_*,\rmi k_*)=0$,  $\partial_\lambda d(\lambda_*,\rmi k_*)\neq 0$;
\item $\lambda_*$ is maximal, $d(\lambda_*+\rho,\rmi k)\neq 0$ for $\rho>0$, $k\in\R$;
\item $\lambda_*$ is locally extremal: the locally unique eigenvalue $\lambda_*(k)$ has $\partial\Re\lambda_*(k)/\partial k=0$.
\end{itemize}
Then $\lambda_*-\rmi s_\mathrm{g}k_*$ is an algebraic pointwise growth mode in a frame with speed $s_\mathrm{g}$, with $\nu_*=\rmi k_*$.
\end{Lemma}
\begin{Proof}
Without loss of generality, $\lambda_*=\nu_*=0$. 
Since $\lambda_*$ is simple, we can locally solve $d(\lambda,\nu)=0$ for $\lambda=\lambda(\nu)$. 

We claim  $\partial_{\Im\nu}\Re\lambda=0$. Notice that passing to the comoving frame shifts the essential spectrum as $\lambda(\rmi k) \mapsto \lambda(\rmi k)-s\rmi k$. Since $\Re\lambda_*$ was extremal (in fact a global maximum) in the steady frame, it is therefore also extremal in the comoving frame, which proves the claim. 

Next, notice that $\partial_{\Im\nu}\Im\lambda=0$ since we passed to a frame where $s_\mathrm{g}=0$. Therefore, $\lambda_*=0$ corresponds to a double root. 

We need to show that the double root is pinched. We therefore choose $p\geq 2$ such that $d(0,\nu)=d_p\nu^p+\rmO(\nu^{p+1})$, and solve for roots $\nu_j(\lambda)=c\lambda^{1/p}\rme^{2\pi\rmi j/p}+\rmO(\lambda^{2/p})$, $j=1,\ldots,p$, using the Newton polygon. Since $\partial_\lambda d(0,0)\neq 0$, we have $c\neq 0$. In particular, there is a root with $\Re\nu>0$ and a root with $\Re\nu<0$ along the curve $\lambda=\tau$, $\tau\geq 0$ when $p>2$ or when $\Re c\neq 0$. The case $\Re c=0$, $p=2$ can be excluded since in this case $d(\lambda,\nu)\sim \lambda+\nu^2$ so that the essential spectrum lies on a curve $\lambda(k)\sim  -\frac{1}{c^2} k^2$, contradicting the assumption that $\lambda=0$ was a maximum of the essential spectrum. 
\end{Proof}

The lemma guarantees that the supremum in the definition of the spreading speed is not taken over the empty set provided that the essential spectrum is unstable. On the other hand, spreading speeds are finite in parabolic equations such as the one considered here.

\begin{Lemma}[Bounded Spreading Speed]\label{l:sinf}
There exists $s_*>0$ so that all pinched double roots lie in $\Re\lambda<0$ for $s>s_*$. 
\end{Lemma}
\begin{Proof}
For $s$ sufficiently large, we claim that $\sigma_\mathrm{ess}^\eta$ for $\eta<0$ sufficiently negative is located in $\Re\lambda<0$. Scaling $s=\hat{s}R^{2m-1}$, $\lambda=\hat{\lambda}R^{2m}$, and $\nu=\hat{\nu}R$, we find that \[
d(\lambda-s\nu,\nu)=R^{2mN}\mathrm{det}\,(A_{2m}\hat{\nu}^{2m}+\hat{s}\hat{\nu}-\hat{\lambda})+\rmO(R^{(2m-1)N}).
\]
Dropping hats, setting $\nu=-1+\rmi k$, and substituting the eigenvalues $\rho$ for the matrix $A$, we find at leading order
\[
\lambda -\rho(-1+\rmi k)^{2m}-s(-1+\rmi k)=0,
\]
so that $\Re\lambda\leq C-s$ for all $k$ after exploiting $(-1)^m\Re\rho<0$.  Introducing lower order terms and going back through the scaling, we find that for speeds $s=s_*R^{2m-1}$, $s_*$ sufficiently large, and weight $\eta=-R$, the real part of the essential spectrum is bounded by $-\lambda_*R^{2m}$ for some $\lambda_*>0$.
\end{Proof}

\begin{Corollary}[Finite Spreading Speed]
Spreading speeds near unstable states are well defined and finite. In a frame moving with $s=s_+$, there is a pinched double root $\lambda=\rmi\omega_*$, $\nu=\nu_*$ located on the imaginary axis. 
\end{Corollary}
\begin{Proof}
Spreading speeds are finite and well-defined by Lemmas \ref{l:sinf} and \ref{l:max}. Continuity of pinched double roots gives the existence of a pinched double root on $\rmi\R$ at $s=s_+$. 
\end{Proof}

\begin{Remark}[Spatial Decay of Invasion Modes $\Re\nu\leq 0$]\label{r:snu}
If the pinched double root on the imaginary axis at $s=s_+$ is simple, we can infer that $\Re\nu_* \leq 0$. In fact, locally near the double root, the dispersion relation has the expansion $\lambda-(s-s_+)\nu_*-a\nu^2+\ldots=0$, which shows that the double root moves as $\lambda\sim (s-s_+)\nu_*$. Since $\Re\lambda\leq 0$ for $s>s_+$, we can conclude $\Re\nu_*\leq 0$. 
\end{Remark}

\begin{Remark}[Spreading intervals]\label{r:gen}
Of course, one can also define spreading speeds  $s_-$ to the left, most easily by reflecting $x\mapsto -x$ and computing $s_+$ in the reflected system. More precise information on the spreading behavior is contained in spreading sets $S$, which are subsets of $\R$ so that there are unstable pinched double roots in $\Re\lambda>0$ for $s\in S$. Continuous dependence of pinched double roots on $s$ shows that the complement of $S$ is open.
\end{Remark}

\begin{Remark}[Reflection Symmetric Systems]\label{r:ref}
Suppose that the system under consideration possesses a reflection symmetry $x\mapsto -x$, possibly combined with an involution $u\mapsto J u$, $J^2=\mathrm{id}$. Equivalently, $A(\partial_x)J=JA(-\partial_x)$, so that $d(\lambda,\nu)=d(\lambda,-\nu)$. We can then write 
\[
d(\lambda,\nu)=:d_1(\lambda,\nu^2).
\]
Of course, in this case $S=-S$, that is propagation to the right and to the left are equivalent. Also, group velocities vanish when $\nu=0$, since $\partial_\nu d=2\nu\partial_\nu d_1$. Also, $\nu\in\rmi\R$ and $\lambda\in\R$ is robust since we can solve $d_1(\lambda,-k^2)$ as a real equation. In this case, group velocities are purely imaginary and automatically vanish when $\lambda(k)$ is extremal. On the other hand, for $\lambda$ complex and $\nu\neq 0$, group velocities do not vanish in general. Such examples arise in local instabilities with nonzero frequency and wavenumber, sometimes referred to as Turing-Hopf; see for instance \cite[\S 2.2]{rad}.
\end{Remark}

\begin{Lemma}[Upper Semi-Continuity]\label{l:usc}
Spreading speeds are upper semi-continuous with respect to system parameters.
\end{Lemma}
\begin{Proof}
Fix a system with spreading speed $s_+^*$. Continuity of pinched double roots and parabolicity imply that for any $\varepsilon>0$ there exists $\delta_1>0$ so that $\Re\lambda_\mathrm{dr}\leq -\delta_1$ for any pinched double root $\lambda_\mathrm{dr}$ and all $s\geq s_+^*+\varepsilon$. Continuity of $\lambda_\mathrm{dr}$ with respect to system parameters and a priori upper bounds on $s$ imply that for systems that are $\delta$-close, $\Re\lambda_\mathrm{dr}\leq -\delta_1/2$ for all  pinched double roots $\lambda_\mathrm{dr}$ and all $s\geq s_+^*+\varepsilon$. This implies that $s_+<s_+^*+\varepsilon$ for all nearby systems, thus establishing upper semi-continuity.
\end{Proof}
\begin{Remark}[Non-continuity]
Consider 
\begin{align*}
u_t&=u_{xx}+u\\
v_t&=v_{xx}+(\mu  +4 \rmi)v - 3v_x.
\end{align*}
The dispersion relation factors,
\[
d(\lambda-s\nu,\nu)=(\nu^2+s \nu+1-\lambda)(\nu^2+(s-3)\nu+\mu+4\rmi-\lambda).
\]
Double roots from the first factor are pinched and stabilize at $s=2$. Double roots from the second factor are also pinched but always stable when $\mu<0$, with nonnegative real part at $s\sim 3$ for $\mu\geq 0$, $\mu\ll 1$. Double roots resulting from collisions of roots from the first and second factor solve $\nu=(\mu-1+4\rmi)/3$, hence yield $\Re\lambda<0$ for $\mu\ll 1$ and $c>0$. In summary, we have that for $\mu<0$, $s_+=2$, but for $\mu>0$, $s_+=3+2\sqrt{\mu}$. Adding an equation $w_t=w_{xx}+(\mu+4\rmi) w + 3w_x$ yields a reflection-symmetric example, $J(u,v,w)=(u,w,v)$. One can also construct examples without gradients exploiting the fact that the group velocity of marginally unstable modes in Turing-Hopf instabilities \cite[\S 2]{rad} is typically non-zero at onset. One can also see from this example that continuity cannot be achieved with small modifications in the definition, such as taking the supremum over speeds where $\Re\lambda_\mathrm{dr}>0$.
\end{Remark}

\section{Spreading in Multi-dimensional Space}
\label{s:spmult}

The spreading of an instability in multi-dimensional space often occurs in the form of roughly radial propagation of disturbances. After initial transients, such behavior can be well described by the unidirectional propagation of a possibly transversely modulated planar interface. One can understand such behavior by studying spreading behavior into a fixed direction, say the $x$-coordinate, of a mode that is extended in the $y$-direction. More precisely, we consider modes that are modulated in the $y$-direction in the form $\rme^{\rmi k_y y}$. For the sake of notation, we restrict to $(x,y)\in\R^2$ and consider the parabolic equation 
\begin{equation}\label{e:00}
u_t=A(\partial_x,\partial_y)u,
\end{equation}
with initial conditions that are localized in $x$, $u(0)=u_0(x)\rme^{\rmi k_y y}$, which gives the parameterized family of equations
\begin{equation}\label{e:11}
u_t=A(\partial_x,\rmi k_y)u.
\end{equation}
We can now repeat the discussion of one-dimensional systems and define spreading speeds $s_+(k_y)$ for each fixed $k_y$. Parabolicity implies that (\ref{e:11}) is stable for $|k_y|$ sufficiently large. Since $s_+$ depends upper semi-continuously on system parameters, Lemma \ref{l:robust}, we can conclude that $s_+(k_y)$ attains its maximum at some finite $k_y$. 

\begin{Definition}[Transverse Modulation]\label{d:trans}
We say that the invasion process is transversely modulated if $s_+$ does not attain its maximum at $k_y=0$. There then exists $k_y\neq 0$ so that $s_+(k_y)\geq s_+(k)$ for all $k$ where $s_+$ is defined, and $s_+(k_y)> s_+(0)$.  We then call $k_y$ a transverse selected wavenumber of the invasion process.  
\end{Definition}

One can easily construct examples where $k_y\neq 0$ in anisotropic systems, considering for instance 
\[
u_t=u_{xx}-(\partial_{yy}+1)^2 + \mu u,
\]
with $0<\mu<1$, where modes with $k_y=0$ are in fact stable. Such invasion processes are often observed when one-dimensional patterns, such as roll solutions in convection experiments, are conquered by hexagon patterns through an invasion process, effectively breaking the transverse $y$-translation symmetry. 

A more interesting question is whether transversely invasion processes occur in isotropic systems, which will be the topic of the remainder of this section.

\subsection{Isotropic Systems}

We consider systems that are isotropic, that is, invariant with respect to rotations and reflections. Again, we restrict to $(x,y)\in\R^2$ for simplicity of exposition, the discussion readily generalizes. 

We say that (\ref{e:00}) is isotropic when $u(t,x,y)$ is a solution if and only if $T(\gamma)u(t,\gamma\cdot(x,y))$ is a solution for any $\gamma\in O(2)$, and $T$ is a representation of $O(2)$ on $\R^N$. Equivalently, $A(\nabla_{(x,y)})=T^{-1}(\gamma)A(\gamma^{-1}\nabla_{(x,y)})T(\gamma)$ for any $\gamma\in O(2)$. The ansatz $u(t,x,y)=\rme^{\lambda t + \nu_x x + \nu_y y} u_0$ now gives the dispersion relation 
\[
d(\lambda,\nu_x,\nu_y)=\mathrm{det}\,(A(\nu_x,\nu_y)-\lambda).
\]
In the isotropic case, the previous discussion implies that $
d(\lambda,\rmi k_x,\rmi k_y)$ only depends on the length of the wave vector $(k_x,k_y)$, so that it can be expressed as a function of $k_x^2+k_y^2$, only. This extends to complex wavenumbers so that
\begin{equation}\label{e:diso}
d(\lambda,\nu_x,\nu_y)=\tilde{d}(\lambda,\nu_x^2+\nu_y^2).
\end{equation}
For simplicity of notation, we will drop tildes in the following and write $d(\lambda,\nu^2-\ell)$, with $\ell=k_y^2$.  Pinched double  roots in $x$-comoving frames solve 
\begin{equation}\label{e:driso}
d(\lambda-s\nu,\nu^2-\ell)=0,\qquad (-s\partial_1+2\nu\partial_2) d(\lambda-s\nu,\nu^2-\ell)=0.
\end{equation}

\subsection{Transverse Pattern Formation}

Our main result shows that there are no transversely modulated invasion processes. 
\begin{Theorem}[Planar Fronts are Fastest]\label{t:1}
Transversely modulated invasion processes do not exist. More precisely, $s_+(k_y)$ attains its maximum at $k_y=0$. 
\end{Theorem}

\begin{Remark}[Transverse Pattern Formation]\label{r:nol}
We stress that the theorem concerns \emph{linear} predictions in the leading edge.  It has indeed frequently been noticed that for all those invasion processes, stripes parallel to the front interface dominate the leading edge of the front; see for instance  \cite{pismen,misbah,hari,pfc}. Of course, nonlinear systems may well exhibit transversely modulated patterns in the wake of a primary invasion. Our point here is that the emergence of transverse modulation is a nonlinear phenomenon, caused by secondary invasion or fast nonlinear, so-called pushed fronts. We refer to the discussion sections for more details. 
\end{Remark}
The proof of this result will occupy the remainder of this section. The key calculation is an implicit differentiation of the dispersion relation which reveals that $s_+$ is strictly decreasing in $k_y^2$. Since, in general, spreading speeds may not be differentiable in the parameter $k_y$, we approximate the dispersion relation by a nearby dispersion relation where this dependence is piecewise smooth. For the approximation, we rely on transversality and perturbation arguments, while keeping the special structure of the dispersion relation dictated by isotropy. 

To be precise, we consider dispersion relations $d_a(\lambda,\nu^2)$, where $a\in\C^M$ denotes the coefficients of the complex multivariable polynomial $d_a$. In this notation, we define 
\[
\mathcal{G}(\omega,s,\nu,\ell,a):=\left(\begin{array}{c}
d_a(\rmi\omega-s\nu,\nu^2-\ell)\\
-s\partial_1 d_a(\rmi\omega-s\nu,\nu^2-\ell)+2\nu \partial_2 d_a(\rmi\omega-s\nu,\nu^2-\ell)
\end{array}\right).
\]
We sometimes write $\partial_\nu:=-s\partial_1+2\nu\partial_2$, $\partial_{\nu\nu}:=s^2\partial_{11}-4s\nu\partial_{12}+4\nu^2\partial_{22}+2\partial_2$,\ldots

We also consider 
\[
\mathcal{G}_{\mathrm{ext},1}(\omega,s,\nu,\ell,a)=\left(\begin{array}{c}
d_a\\
\partial_\nu d_a\\
\partial_\lambda d_a
\end{array}
\right),\qquad 
\mathcal{G}_{\mathrm{ext},2}(\omega,s,\nu,\ell,a)=\left(\begin{array}{c}
d_a\\
\partial_\nu d_a\\
\partial_{\nu\nu} d_a
\end{array}
\right).
\]
Note that $\mathcal{G}_{\mathrm{ext},1}$ or $\mathcal{G}_{\mathrm{ext},2}$ vanish precisely at multiple double roots. 

We introduce coefficients of $d_a$ explicitly via the expansion at the origin,
\[
d_a(\lambda,\nu^2)=a_0+a_{10}\lambda+a_{01}\nu^2+a_{02}\nu^4+\ldots
\]
We can now calculate derivatives of $\mathcal{G}_{\mathrm{ext},j}$ with respect to those coefficients:
\begin{align}\label{e:der}
\partial_{a_0}\mathcal{G}_{\mathrm{ext},1}&=
\left(\begin{array}{c}
1\\0\\0
\end{array}\right),\qquad
\partial_{a_{10}}\mathcal{G}_{\mathrm{ext},1}&=
\left(\begin{array}{c}
\rmi\omega -s\nu\\-s \\1
\end{array}\right),
\qquad
\partial_{a_{01}}\mathcal{G}_{\mathrm{ext},1}&=
\left(\begin{array}{c}
\nu^2-\ell\\2\nu\\0
\end{array}\right),
\nonumber\\
\partial_{a_0}\mathcal{G}_{\mathrm{ext},2}&=
\left(\begin{array}{c}
1\\0\\0
\end{array}\right),
\qquad
\partial_{a_{01}}\mathcal{G}_{\mathrm{ext},2}&=
\left(\begin{array}{c}
\nu^2-\ell\\2\nu\\2
\end{array}\right),
\qquad
\partial_{a_{02}}\mathcal{G}_{\mathrm{ext},2}&=
\left(\begin{array}{c}
(\nu^2-\ell)^2\\4\nu(\nu^2-\ell)\\12\nu^2-4\ell
\end{array}\right).
\end{align}
Consider now the domain of $\mathcal{G}$ excluding $\nu=0$, 
\begin{equation}\label{l:eg}
\mathcal{V}=\{\omega\in\R,s\in\R,\nu\in\C\setminus\{0\},\ell\in\R,a\in\C^M\}\subset \R^2\times\C\times\R\times\C^M\sim \R^4\times\R\times\C^M,
\end{equation}
so that 
\begin{equation}\label{e:g}
\mathcal{G}:\mathcal{V}\to \C^2\sim \R^4,\qquad \mathcal{G}_{\mathrm{ext},j}:\mathcal{V}\to \C^3\sim\R^6.
\end{equation}
Our goal is to move the parameters $a$ to ensure that the $\mathcal{G}_{\mathrm{ext},j}$ do not vanish. We will accomplish this by using transversality. We adopt the usual definition, where a smooth map between smooth manifolds $\mathcal{H}:\mathcal{U}\to\mathcal{W}$ is transverse to a smooth  submanifold $\mathcal{Z}$ of $\mathcal{W}$ if $\Rg(D\mathcal{H}(u))+T_{\mathcal{H}(u)}\mathcal{Z}=T_{\mathcal{H}(u)}\mathcal{W}$ for all $\mathcal{H}(u)\in\mathcal{Z}$. In our specific case $\mathcal{Z}=\{0\}$ is a point, and $\mathcal{U}$ and $\mathcal{W}$ are open subsets of $\R^{j}$ and $\R^J$, respectively. Transversality is then equivalent to the fact that the derivative is onto. 
\begin{Lemma}\label{l:t0}
The maps $\mathcal{G}_{\mathrm{ext},1}$ and $\mathcal{G}_{\mathrm{ext},2}$, considered on domains defined in (\ref{e:g}), are transverse to $\{0\}$. More specifically, $\partial_a \mathcal{G}_{\mathrm{ext},1}$ and $\partial_a \mathcal{G}_{\mathrm{ext},2}$
are onto. 
\end{Lemma}
\begin{Proof}
Inspecting the formulas for partial derivatives (\ref{e:der}) shows that $\partial_{a_{j}}\mathcal{G}_{\mathrm{ext},1}$, $j\in \{1,10,01\}$ are linearly independent over $\C$ as long as $\nu\neq 0$, which establishes that the range of $\partial_a \mathcal{G}_{\mathrm{ext},1}$ is real 6-dimensional. This implies transversality of $\mathcal{G}_{\mathrm{ext},1}$ to $\{0\}$. Similarly,  $\partial_{a_{j}}\mathcal{G}_{\mathrm{ext},2}$, $j\in \{1,01,02\}$ are linearly independent over $\C$ and $\mathcal{G}_{\mathrm{ext},2}$ is transverse to $\{0\}$.
\end{Proof}
Using Sard's transversality theorem \cite{AR}, we can conclude that the restriction to a fixed parameter $\mathcal{G}_{\mathrm{ext},j}^a(\cdot):=\mathcal{G}_{\mathrm{ext},j}(\cdot,a)$ is transverse for a residual, in particular dense, subset of parameter values.
\begin{Corollary}\label{c:t0}
For all $a$ in a residual subset $\mathcal{S}_0\subset\C^M$, 
\[
\mathcal{G}_{\mathrm{ext},1}^a\neq 0,\qquad 
\mathcal{G}_{\mathrm{ext},2}^a\neq 0,\qquad \mbox{ for all } \omega,s,\ell\in\R,\nu\in\C\setminus \{0\}.
\]
\end{Corollary}
\begin{Proof}
Sard's transversality implies that $\mathcal{G}_{\mathrm{ext},j}^a$ are transverse to $\{0\}$ for $a$ in a residual subset of $\C^M$. Since the domain is 5-dimensional and the target manifold is 6-dimensional, the linearization cannot be onto, hence transversality implies that $0$ is not in the image. 
\end{Proof}
We will use a very similar transversality argument to exclude $\nu\in\rmi\R$ at double roots, provided that $\ell=0$. Consider therefore 
\[
\tilde{\mathcal{G}}(\omega,s,k,a):=\mathcal{G}(\omega,s,\rmi k,0,a)
\]
on 
\[
\tilde{\mathcal{V}}=\{\omega\in\R,(k,s)\in\R^2\setminus\{0\},a\in\C^M\}
\]
We also define analogously maps $\tilde{\mathcal{G}}^a$ via restriction. 
\begin{Lemma}\label{l:t1}
The map $\tilde{\mathcal{G}}$ considered on $\tilde{\mathcal{V}}$, is transverse to $\{0\}$. More specifically, $\partial_a \tilde{\mathcal{G}}$ is onto.
\end{Lemma}
\begin{Proof}
For $s\neq 0$, $\partial_{a_0}\tilde{\mathcal{G}}=(1,0)^T$ and $ \partial_{a_{10}}\tilde{\mathcal{G}}=(*,-s)^T$ are linearly independent over $\C$. For $\nu\neq 0$, $\partial_{a_0}\tilde{\mathcal{G}}$ and  $ \partial_{a_{01}}\tilde{\mathcal{G}}=(*,2\rmi k)^T$ are linearly independent.
\end{Proof}
\begin{Corollary}\label{c:t1}
For all $a$ in a residual subset $\mathcal{S}_1\subset\mathcal{S}_0$, 
\[
\tilde{\mathcal{G}}^a\neq 0,\qquad \mbox{ for all } \omega\in\R,(s,k)\in\R^2\setminus\{0\}.
\]
\end{Corollary}
\begin{Proof}
Again, we conclude from Sard's transversality that $\tilde{\mathcal{G}}^a$ is transverse in a residual subset. Since the target space is (real) 4-dimensional, the domain only 3-dimensional, transversality implies that there are no roots of $\tilde{\mathcal{G}}^a$.
\end{Proof}
We say that a solution of $\mathcal{G}^a(\omega,s,\nu,\ell)=0$ is a simple spreading speed if $\partial_{(\omega,s,\nu)}\mathcal{G}^a$ is invertible at the solution. 
\begin{Proposition}\label{p:t}
For all $a\in\mathcal{S}_1$, spreading speeds are simple unless $\nu=s=0$. 
\end{Proposition}
\begin{Proof}
We compute 
\[
\partial_{(\omega,s,\nu)}\mathcal{G}^a=\left(\begin{array}{ccc}
\rmi\partial_1d & -\nu\partial_1 d & 0\\
* & * & \partial_{\nu\nu}d
\end{array}\right).
\]
Here, the first two columns are acting on $\R\times \R$ and the last column on $\C$. Note that, considered as a map on $\R^4$, this matrix is invertible provided $\nu\not\in\rmi\R$ since $\partial_1 d$ and $\partial_{\nu\nu} d$ do not vanish at solutions for $a\in\mathcal{S}_0\supset \mathcal{S}_1$, Corollary \ref{c:t0}. We next claim that $0\neq \nu\in\rmi\R$ is not possible for a spreading speed when $a\in\mathcal{S}_1$. Note that Corollary \ref{c:t1} guarantees this fact in the case $\ell=0$, only. Suppose therefore that $\omega_*,s_*,\nu_*=\rmi k_*,\ell_*$ are a root of $\mathcal{G}^a$. Using isotropy, one directly verifies that another solution is given by 
\[
\tilde{k}^2:=k_*^2+\ell_*,\quad \tilde{s}=2\rmi\tilde{k}\frac{\partial_2 d}{\partial_1d},\quad \tilde{\omega}=\tilde{s}\tilde{k}.
\]
In the substitution, one exploits that 
$\rmi\tilde{\omega}-\tilde{s}\rmi\tilde{k}=\rmi\omega_* -s_*\rmi k_*$, so that the arguments of $d$ remain the same upon substitution. Note also that $\frac{\partial_2 d}{\partial_1 d}\in\rmi\R$ since $\partial_\nu d=0$. Summarizing, we have found a solution $\mathcal{G}^a(\tilde{\omega},\tilde{s},\tilde{k},0)=\tilde{\mathcal{G}}^a(\tilde{\omega},\tilde{s},\tilde{k})=0$, which however was excluded by Corollary \ref{c:t1}, with the exception of the case $\nu=s=0$. This proves the Lemma.
\end{Proof}
As a consequence, choosing $a\in\mathcal{S}_1$, we find that solutions to $\mathcal{G}^a(\omega,s,\nu,\ell)=0$ come as smooth curves $(\omega,s,\nu)(\ell)$, with end points (and possible singularities) only at $\nu=s=0$. Also, $\Re\nu\neq 0$ on these curves unless $\nu=s=0$. 

\begin{Lemma}[Monotonicity]\label{l:mon}
Suppose $a\in\mathcal{S}_1$ and let $(s,\omega,\nu)(\ell)$ be a generalized spreading speed with $s>0$. Then 
\begin{equation}\label{e:d12}
\frac{\rmd (s^2)}{\rmd \ell}=-\frac{s^2}{|\nu|^2}<0.
\end{equation}
\end{Lemma}
\begin{Proof}
Recall that 
\begin{equation}\label{e:fu}
\frac{\partial_2 d}{\partial_1 d}=\frac{s}{2\nu}.
\end{equation}
Expanding  $\mathcal{G}^a$ near a solution and denoting by $\hat{\omega}, \hat{s},$ and $\hat{\ell}$ the increments, we find at first order
\[
\rmi (\partial_1 d)\hat{\omega}-\nu(\partial_1 d)\hat{s}-(\partial_2 d)\hat{\ell}=0.
\]
Exploiting (\ref{e:fu}) we find 
\[
\rmi\hat{\omega}-\nu \hat{s}=\frac{s}{2\nu}\hat{\ell},
\]
and, taking real parts, 
\[
\hat{s}=-\frac{\Re\frac{s}{2\nu}}{\Re\nu}\cdot\hat{\ell}=-\frac{\Re\frac{s\bar{\nu}}{2|\nu|^2}}{\Re\nu}\cdot\hat{\ell}.
\]
Differentiating gives $\frac{\rmd s}{\rmd \ell}=\frac{\hat{s}}{\hat{\ell}}$ and the desired result. 
\end{Proof}

\begin{Proof}[of Theorem \ref{t:1}]
We argue by contradiction.  Consider a dispersion relation $d$, associated polynomial coefficients $a$, so that $s_+(k_y)>s_+(0)$ for some $k_y$. 
We would like to consider systems with $\hat{a}\in \mathcal{S}_1$. Therefore, first modify the dispersion relation setting $\tilde{d}(\lambda,\nu^2)=d(\lambda-\varepsilon_1,\nu^2)$, for some $\varepsilon_1>0$ sufficiently small, and write $\tilde{a}$ for the associated vector of coefficients. Since this perturbation merely shifts values of $\lambda$, double roots are simply shifted by $\varepsilon_1$. Now choose $\hat{a}\in \mathcal{S}_1$ $\varepsilon_2$-close to $\tilde{a}$. By continuity of pinched double roots, the real part of pinched double roots for $\hat{a}$ will be strictly larger than the real part of double roots for $a$ as long as $\varepsilon_2\ll \varepsilon_1$. As a consequence, the associated spreading speeds $s_+(k_y)$ and $\hat{s}_+(k_y)$ satisfy $s_+(k_y)< \hat{s}_+(k_y)$. Using upper semi-continuity, Lemma \ref{l:usc}, we conclude that $\hat{s}_+(k_y)- s_+(k_y)<\varepsilon$, arbitrarily small provided $\varepsilon_1,\varepsilon_2$ are sufficiently small. In particular, $\hat{s}_+(k_y)>\hat{
s}_+(0)$

Since the spreading speed $\hat{s}_+$ is realized by a finite number of pinched double roots on the imaginary axis, all of which satisfy the monotonicity formula from Lemma \ref{l:mon}, the spreading speed is strictly decreasing for each $\ell=k_y^2>0$. This contradicts our assumption and proves the theorem.
\end{Proof}

\section{Summary and Discussion}\label{s:dis}

We summarize our results, Section \ref{s:sum}, and comment on systems without translation symmetry in Section \ref{s:inh}. We then comment extensively on challenges with nonlinear systems, Section \ref{s:nonl}, and conclude with a short outlook in Section \ref{s:con}.

\subsection{Summary}\label{s:sum}

We considered generalized spectral indicators for pointwise growth and associated growth rates. For linear systems, pointwise growth modes (PGM) determine exponential decay and growth in a finite window of observation for a system on the real line. Pointwise growth modes correspond to singularities of pointwise projections $P^\mathrm{s}_\lambda$. 

When the domain is the positive half line, right-sided pointwise growth modes (RPGM) take this role, at least for suitable boundary conditions. Right-sided pointwise growth modes correspond to singularities of the stable subspace $E^\mathrm{s}_\lambda$ and are a subset of pointwise growth modes. 

Pinched double roots (PDR) are defined via determinants rather than matrices and determine pointwise growth only in generic situations. As opposed to (one-sided) pointwise growth modes, they are however continuous with respect to system parameters. From an algorithmic point of view, one can compute double roots (DR), then specialize to pinched double roots, and finally check on the presence of pointwise and right-sided pointwise growth modes; we refer to \cite{drcomp,rss} for computational aspects of the first steps in this procedure.  

We gave a number of examples that highlight the difference between these concepts. A key role was played by the example of counter-propagating waves (CPW),
\begin{align*}u_t&=u_{xx}+u_x\\v_t&=v_{xx}-v_x+\mu u.\end{align*} 
The following table summarizes some of our results, listing existence of growth modes or pinched double roots at $\lambda=0$ in the examples, as well as continuity, semi-continuity, and availability (and continuity)  of counts.
\begin{center}
\begin{tabular}{ccccccc}
\toprule 
& $u_t=u_{xx}$ & CPW $\mu\neq 0$ & CPW, $\mu=0$ & cont. & lower semi-cont. & counts \\
\midrule
DR & yes & yes& yes& yes & yes& yes\\
PDR & yes & yes& yes& yes & yes& no\\
PGM& yes & yes& no & no& yes& no\\
RPGM& yes & no& no & no& yes& no\\
\bottomrule
\end{tabular}
\end{center}
In most examples that we have encountered, double roots appear to be most amenable to explicit analysis. The pinching condition can be more cumbersome to analyze. Pointwise growth modes and right-sided pointwise growth modes need only be computed in the non-generic cases when multiple double roots determine growth. In such cases, one can focus on a \emph{local} analysis near the pinched double root and compute $P^\mathrm{s}_\lambda$ or $E^\mathrm{s}_\lambda$, which can then often be split in singular and non-singular subspaces $E^\mathrm{s}(\lambda)=E^\mathrm{ss}(\lambda)\oplus E^\mathrm{c}(\lambda)$, $E^\mathrm{ss}(\lambda)$ analytic. 

Based on pinched double roots, we defined spreading speeds as maximal speeds of comoving frames with marginally stable pinched double roots. We do not know if pointwise growth modes or right-sided pointwise growth modes are continuous with respect to changes in the laboratory frame. As a consequence, a definition of spreading speeds based on these more subtle concepts would be less workable at this point. 

As an application, we studied linear spreading speeds in two-dimensional domains, depending on a transverse wavenumber. We showed that linearly determined, transversely planar, non-modulated fronts are always fastest. We do not have a simple intuitive explanation of this fact.

\subsection{Inhomogeneous Linear Systems and Resonance Poles}\label{s:inh}
We discuss generalizations and new phenomena associated with spatially inhomogeneous systems, 
\[
u_t = A(\partial_x,x)u, \quad u\in\R^N, \ x\in\R,
\]
where $A$ is smoothly depending on $x$ and ellipticity conditions (\ref{e:ell}) are satisfied uniformly in $x$. We discuss periodic and homoclinic/heteroclinic coefficients. 

\paragraph{Periodic Coefficients.}
In $L$-periodic media, one can follow the exposition in this paper very closely and construct pointwise first-order Green's functions using the $x$-periodic linear evolution $\Phi_\lambda(\xi,\zeta)$ to the first-order equation $U_x=M_\lambda(x) U$. Analyticity properties of the Green's function $T_\lambda(\xi,\zeta)=T_\lambda(\xi+L,\zeta+L)$ are independent of $\xi,\zeta$. They depend only on the pointwise projection $P^\mathrm{s}_\lambda(\xi)=P^\mathrm{s}_\lambda(\xi+L)$. This can be readily seen using Floquet theory, which transforms the $x$-periodic linear differential equation into a constant-coefficient system via an $x$-dependent change of variables. One can also define an analytic dispersion relation via 
\[
d(\lambda,\nu)=\mathrm{det}\,(\Phi_\lambda(L,0)-\rme^{\nu L}).
\]
Continuity results carry over, but counts do not apply since the dispersion relation is not polynomial. In fact, there are typically infinitely many double roots. We refer to \cite{BB} for a discussion of pointwise growth and double roots in this context.  

Periodic coefficients arise for instance when studying secondary invasion. As we saw in Section \ref{s:spmult}, primary pattern-forming  invasion typically creates one-dimensional stripes parallel to the front interface. Often these striped patterns are unstable and a secondary invasion process will create more complex patterns such as squares and hexagons. This secondary invasion process can to some approximation be studied using the linearization at the primary, unstable striped pattern. Of course, the linearization at this striped pattern will not be isotropic, even if the underlying equation is, so that one may now observe transversely modulated fronts. 

Beyond periodic coefficients, generalizations to quasi-periodic and random media have been studied, mostly in scalar equations. We refer to \cite{quasi,quasi2,random} and references therein without attempting a generalization of our concepts in this direction.

\paragraph{Asymptotically Constant Coefficients.}
Also of interest are situations where $A(\partial_x,x)\to A_\pm(\partial_x)$ is heteroclinic (or homoclinic when $A_+=A_-$). To some extent, this case has been studied extensively in the context of stability problems of nonlinear waves using the Evans function; see for instance \cite{kp,san}. The relation with pointwise growth becomes apparent in this context when extending Evans functions across the essential spectrum using the Gap Lemma. In the context of pointwise stability, our discussion here is similar to \cite{brevdohet}.

Associated with $A_+$, we consider the subspaces $E^\mathrm{s}_\lambda$, associated with $A_-$ we consider $E^\mathrm{u}_\lambda$. For $\lambda\gg 1$, these subspaces contain bounded solutions on $x>0$ and $x<0$, respectively, to the first-order equations $U_x=M^\pm_\lambda U$ associated with $A_\pm$. 

Assuming sufficiently rapid convergence in $x$\footnote{That is, with sufficiently strong exponential rate; see \cite{SSblowup} for cases when convergence is too weak.}, the $x$-dependent problem possesses subspaces $E^\mathrm{s/u}_\lambda(x)$ that contain initial conditions to bounded solutions on $x>0$ and $x<0$, respectively, for the $x$-dependent problem $U_x=M_\lambda(x)U$. Moreover, these subspaces differ from $E^\mathrm{s/u}_\lambda$ by an analytic linear transformation, only.  The first-order Green's function is given by 
\begin{equation}\label{e:genon}
T_\lambda(x,y) = \begin{cases}
\Phi_\lambda(x,y)P^\mathrm{s}_\lambda(y) , & x>y , \\
-\Phi_\lambda(x,y)P^\mathrm{u}_\lambda(y) , & x<y,
\end{cases}
\end{equation}
where $P^\mathrm{s/u}_\lambda(y)$ are the projections along $E^\mathrm{u/s}_\lambda(y)$ onto $E^\mathrm{s/u}_\lambda(y)$. Singularities of the Green's function therefore stem from either
\begin{itemize}
\item singularities of the asymptotic subspaces, in other words, left- and right-sided growth modes, or from
\item intersections between $E^\mathrm{s}_\lambda(y)$ and $E^\mathrm{u}_\lambda(y)$. 
\end{itemize}
The intersections occur in similar fashions as pointwise growth modes or boundary pointwise growth modes and can be tracked using Evans functions. Associated with such intersections are solutions with certain exponential asymptotic behavior $\rme^{\lambda^\pm t + \nu^\pm x}$, that yield spreading speeds via $s=-\Re\lambda/\Re\nu$; see \cite{brevdohet}.

Of course, this discussion can now be combined with the case of periodic coefficients, thus giving a systematic basis to pointwise growth and invasion speeds in problems with \emph{asymptotically periodic coefficients}. We will come back to these issues when discussing nonlinear invasion problems in the next section.

\subsection{Nonlinear Systems}
\label{s:nonl}

We think of the linear theory as a \emph{predictor} for nonlinear phenomena. In the case of simple roots, there are typically open regions in parameter space where linear predictions are correct. We comment below on mechanisms that lead to deviations from linear predictions. 

\paragraph{Simple Growth Modes --- Pushed Fronts.}

For simple pinched double roots, all concepts of pointwise stability studied here coincide, and there is a fairly universal description of associated phenomena \cite{vS}. As far as the invasion speed is concerned, one observes a dichotomy between fronts that propagate with the linear spreading speed (pulled fronts) and fronts that propagate faster than the linear spreading speed (pushed fronts). The prototypical example are fronts in the Nagumo equation 
\[
u_t=u_{xx}+u(1-u)(u-a),
\]
invading the unstable state $u\equiv a$ and leaving behind the stable state $u=1$. For $1/3<a<1/2$, these fronts propagate with the linear speed, for $0<a<1/3$, the invasion speed is faster. More general (explicit) examples are known for the quintic-cubic Ginzburg-Landau equation \cite{vSH}.

In this regard, our analysis here, and the discussion in the sequel, is aimed at pulled fronts, which, loosely speaking, arise when the nonlinearity is not strongly amplifying growth.\footnote{The situation is analogous to supercritical and subcritical bifurcation scenarios: the linearization often gives good predictions in supercritical bifurcations but one does not expect accurate linear predictions in subcritical bifurcations.}

\paragraph{Simple Growth Modes --- Frequencies and Wavenumbers.}

While speed predictions are fairly reliable, wavenumber predictions involve a wider variety of phenomena, even for pulled fronts. We assume that the spreading speed is realized by a simple pinched double root $(\rmi\omega_*,\nu_*)$, which predicts marginal stability in a frame moving with the spreading speed. In other words, we expect to see linear oscillations with frequency $\omega_*$ in this frame of reference. The simplest prediction for patterns in the wake of the front would be to ask for the pattern to be in strong resonance with this frequency, in the comoving frame, so that there would exist a \emph{coherent invasion front} $u(x-s_+t,\omega_\mathrm{f} t)$, $u(\xi,\tau)=u(\xi,\tau+2\pi)$, and $\omega_*= \omega_\mathrm{f}$. This strong resonance is sometimes referred to as ``node conservation'', referring to the actual process of creating patterns with nodes (zeros) which mark the minimal period of the pattern. However,  subharmonic invasion fronts  $\omega_\mathrm{f}=\omega_*/\ell$, $\ell=2,3,\ldots$,
 are also frequently observed, \cite{chfront,reu2011,reu2012}. 

The frequency $\omega_\mathrm{f}$ of the coherent invasion front puts constraints on patterns in the wake of the front. Assume that a wave train is created in the wake of the front, that is,
$|u_\mathrm{f}(x-s_+t,\omega_\mathrm{f}t)- u_\mathrm{wt}(k_-x-\omega_- t;k_-)|\to 0$ for $x\to -\infty$, where $u_\mathrm{wt}(\xi;k)=u_\mathrm{wt}(\xi+2\pi;k)$, and $\omega_-=\omega_-(k_-)$ is the nonlinear dispersion relation in the wake. Periodicity in the comoving frame then requires that 
\begin{equation}\label{e:ks}
\omega_-(k_-)-k_- s_+=\omega_*,
\end{equation}
which, considered as an equation for $k_-$, determines the wavenumber in the wake. Examples are systems such as the Cahn-Hilliard equation or the Swift-Hohenberg equation, with $\omega_-(k)\equiv 0$, which gives $|k_-|=\omega_*/s_+$, as well as the complex Ginzburg-Landau equation 
\[
A_t=(1+\rmi\alpha)A_{xx}+A-(1+\rmi\gamma)A|A|^2,\]
where 
\[
\omega = (\gamma-\alpha) k^2 -\gamma,\quad s_+=2\sqrt{1+\alpha^2},\quad k_-=-\frac{\sqrt{1+\gamma^2}-\sqrt{1+\alpha^2}}{\gamma-\alpha}.
\]
One can sometimes show the existence of such coherent invasion fronts \cite{ce,gms,tc,chfront}, $\omega_\mathrm{f}=\omega_*$, and prove local stability. Selection of slowest fronts however has not been shown in any such context, which makes mathematically rigorous statements on wavenumber selection impossible. Nevertheless, it appears that stability of such a coherent invasion front implies ``node conservation'' in the invasion process, while instabilities lead to changed wavenumbers \cite{chfront}. Note that when referring to stability of a coherent invasion front, we are asking about pointwise stability in the sense discussed in Section \ref{s:inh} with asymptotically periodic coefficients. 

On the other hand, coherent invasion fronts with $\omega=\omega_*$ may simply not exist. A prototypical example are relaxation oscillators of the form 
\begin{align*}
u_t&=u_{xx}+u(1-u)(u+1)-\gamma v,\\
v_t&=v_{xx}+\varepsilon(u-v),
\end{align*}
with $\gamma<1$. 
For $\varepsilon$ small, the equilibrium $u=v=0$ is unstable with selected speed and frequency $s_+\sim 2$, $\omega_*=0$, since the problem is a small perturbation of the scalar $u$-problem. Frequency $0$  would predict a stable stationary pattern in the wake of the front, which however does not exist for the given choice of $\gamma$, for $\varepsilon$ sufficiently small. Stable patterns in the problem are rather modulations of the relaxation oscillation $(u,v)(\omega t)$, with $\omega\sim \varepsilon$. Strongly resonant wavenumber selection (or node conservation) (\ref{e:ks}) then implies $\omega_-(k_-)-k_- s_+=0$, which  implies $k_-\not\sim 0$. One numerically observes phase slips (failure of node conservation) in the leading edge, but this phenomenon does not appear to be well understood theoretically.

\paragraph{Simple Growth Modes --- Secondary fronts and Wavenumber Corrections.}

When a strongly resonant primary front $\omega_\mathrm{f}=\omega_*$ is unstable, we can attempt to predict secondary invasion speeds, frequencies, and wavenumbers based on the linearization at this primary front. Such spreading speeds can now be determined by singularities of the Evans function (resonance poles) or by right-sided pointwise growth modes. Whenever these secondary spreading speeds are slower than the primary speeds, one can expect to see an increasingly long transient of the primary unstable pattern in the growing region between primary and secondary front. When the secondary speed exceeds the primary speed, the secondary front locks to the first front and we immediately see the pattern created by the secondary front, which amounts to an effective correction of the observed wavenumber. 

In \cite{chfront}, secondary spreading speeds and selected wavenumbers were computed based on right-sided pointwise growth modes. It was found that right-sided pointwise growth modes underestimate the secondary invasion speeds, hinting at a resonance pole as the cause of destabilization of the primary front. In \cite{reu2011}, the pattern selected by the primary front is in fact stable, yet we observe locked secondary fronts in certain parameter regimes. 

Beyond these predictions, the phenomenon of staged invasion was investigated theoretically in \cite{HS}, in the context of a simple coupled-mode problem. The predicted secondary spreading speed is based on right-sided pointwise growth modes or resonance poles.  These predictions are validated by the construction of sub- and super-solutions and expansions for the width of the region occupied by the primary pattern in the locked regime are given.

\paragraph{Simple Growth Modes --- Transverse Patterning.}

A similar perspective can also shed light on the formation of transverse patterns through invasion processes. Our results in Section \ref{s:spmult} predict that the primary invasion mechanism creates a striped pattern. We can therefore first restrict to invasion fronts that are independent of $y$, propagating in the $x$-direction, and find fronts
as described above. When studying secondary invasion, however, we need to take into account the possibility of transverse patterning. We can, in principle, repeat the analysis outlined in the one-dimensional case for fixed transverse modulation, $\rme^{\rmi k_y y}$, and study instabilities via right-sided pointwise growth modes or resonance poles. The wavenumber $k_y$ with the fastest spreading speed is then the linear prediction for secondary patterns. Since right-sided pointwise growth modes are evaluated using Floquet theory, transverse patterns can, in principle, be modulated both in $x$ and $y$, and a detailed analysis should distinguish between hexagons and squares, say, in the wake of fronts, as observed in \cite{fw2}, for instance.  

We note at this point that the description of secondary, transversely patterned fronts leads to systems with an invariant subspace given by the $y$-independent solutions. In the simplest context, this is apparent in an amplitude approximation to hexagon-roll competition in the Swift-Hohenberg equation \cite{pismen,misbah,hari,HS}. We therefore expect double double roots similar to the example of counter-propagating waves to occur in a robust fashion. We will discuss nonlinear phenomena associated with double double roots, next. 

\paragraph{Relevant and Irrelevant Double Double Roots.}

With robust examples in ecology \cite{weinbergeranomalous} and pattern formation \cite{pismen,misbah,hari}, double double roots are one of the main challenges that we isolated here. Like any other algebraic pointwise growth mode, these double double roots give linear predictions for the selected speed of the nonlinear system.  In the context of a Lotka-Volterra competition model, a double double root was found that overestimates the invasion speed of the nonlinear system, see \cite{lotka}.  Examples in \cite{hol} show that double double roots sometimes give correct predictions for spreading speeds. In the following, we relate some of the results and observations in \cite{hol} to our point of view.  We will refer to double double roots as {\em relevant} if the linearly selected speed is the nonlinear speed and {\em irrelevant} if the nonlinear speed is slower.  

We will first lay out some general systems of equations that may give rise to double double roots.  We will then relate these double double roots to the concepts of PGMs, RPGMs and BPGMs developed earlier in this article.  An important difference between relevant and irrelevant double double roots will be explained.

We consider the skew-coupled system for $u_1\in \R^{N_1}$, $u_2\in\R^{N_2}$,
\begin{align}
u_{1,t}&=\mathcal{P}_{1}(\partial_x,u_1)u_1\nonumber\\
u_{2,t}&=\mathcal{P}_{2}(\partial_x,u_1,u_2),\label{e:skp}
\end{align}
with appropriate conditions on the nonlinear functions $\mathcal{P}_j$. 
Note that the subspace $u_1=0$ is invariant, but that the skew-product structure of the nonlinear system is not enforced by the presence of this invariant subspace.  We have already encountered one such system in the example of counter propagating waves.
The linearization of (\ref{e:skp}) at the origin possesses a lower block-triangular form,
\begin{align*}
u_{1,t}&=\mathcal{A}_{11}(\partial_x)u_1\\
u_{2,t}&=\mathcal{A}_{21}(\partial_x)u_1+\mathcal{A}_{22}(\partial_x)u_2,
\end{align*}
where 
\[
\mathcal{A}_{11}(\partial_x)=\mathcal{P}_{1}(\partial_x,0),\quad 
\mathcal{A}_{21}(\partial_x)=\partial_{u_1}\mathcal{P}_{2}(\partial_x,0,0),\quad
\mathcal{A}_{22}(\partial_x)=\partial_{u_2}\mathcal{P}_{2}(\partial_x,0),
\]
are differential operators of order $2m$. Due to the skew-product structure, the dispersion relation factors,
\[
d(\lambda,\nu)=d_1(\lambda,\nu)d_2(\lambda,\nu)=\mathrm{det}(\mathcal{A}_{11}(\nu)-\lambda)\,\mathrm{det}(\mathcal{A}_{22}(\nu)-\lambda).
\]
Double double roots now occur in a robust fashion whenever 
\[
d_1(\lambda_*,\nu_*)=d_2(\lambda_*,\nu_*)=0, \qquad  \partial_id_j(\lambda_*,\nu_*)\neq0,\quad i,j\in\{1,2\}.
\]
Indeed, roots $\nu$ are analytic in $\lambda$ as solutions to $d_j(\lambda,\nu)=0$. If the double double root is pinched, then by definition $\lambda_*$ is an algebraic pointwise growth mode.  However, the stable and unstable eigenspaces will remain analytic in a neighborhood of $\lambda_*$ and the double root does not yield a right- (or left-) sided pointwise growth mode. Linearizing the eigenvalue problem in $\nu$ by writing the system 
\begin{align*}
\lambda u_{1}&=\mathcal{A}_{11}(\partial_x)u_1\\
\lambda u_{2}&=\mathcal{A}_{21}(\partial_x)u_1+\mathcal{A}_{22}(\partial_x)u_2,
\end{align*}
as a first-order system in $x$,
\[
\begin{array}{llll}
\mathcal{T}_{11}(\lambda)U_1&-U_{1,x}&=0\\
\mathcal{T}_{21}(\lambda)U_1+\mathcal{T}_{22}(\lambda)U_2&-U_{2,x}&=0,
\end{array}.
\]
The double double root corresponds to eigenvectors $\mathcal{T}_{11}(\lambda_*)e_1=\nu_* e_1$, $\mathcal{T}_{22}(\lambda_*)e_2=\nu_* e_2$. Typically, $\mathcal{T}_{21}(\lambda_*)e_1\not\in\Rg(\mathcal{T}_{22}(\lambda_*)-\nu_*)$, so that the eigenspace to $\nu_*$ in the full system is only one-dimensional and spanned by $(0,e_2)^T$. Once again, this is in complete analogy to the case of the counter-propagating wave problem. We can continue eigenvalues to $\nu_j(\lambda)$ to $\mathcal{T}_{jj}(\lambda)$ analytically and  distinguish three cases:
\begin{enumerate}
\item \emph{uncoupled:} $\mathcal{T}_{21}(\lambda_*)e_1\in\Rg(\mathcal{T}_{22}(\lambda_*)-\nu_*)$;
\item \emph{stable flip}: $\Re\nu_1(\lambda)\to -\infty$ for $\lambda\to\infty$;
\item \emph{unstable flip}: $\Re\nu_1(\lambda)\to +\infty$ for $\lambda\to\infty$.
\end{enumerate}
Case (i) does not give a pointwise growth mode and none of the three cases gives either left- or right-sided pointwise growth modes. 

One does however observe a significant difference between (ii) and (iii) in the context of a system of coupled Fisher-KPP equations, see \cite{hol}.  There, it is observed numerically that if the double double root is of the form (ii), then the double double root is relevant and the nonlinear speed is the linear speed.  On the other hand, if the double double root is of the form (iii) the double double root is irrelevant and the observed speed is slower.  We will now motivate these observations.  Suppose we are considering invasion to the right, with positive spreading speed. Suppose that this invasion occurs as a traveling front moving with a speed $s_0$ that is smaller than the linear invasion speed given by the double double root, and consider the linearization at the associated traveling wave, with associated Evans function. 

In case (ii), the stable subspace at $x=+\infty$ flips at $\lambda_*$, so that its projection on the $U_1$-component is $mN_1-1$-dimensional.
Since the full linearization leaves the $U_1$-subspace invariant, the unstable subspace at $x=-\infty$ is $mN_1$-dimensional, and, again as a consequence of the skew-product structure, stable and unstable subspaces at $x=0$ cannot be transverse, which implies the existence of a resonance pole at $\lambda=\lambda_*$ and pointwise instability of the slower traveling front.  

In case (iii), the unstable subspace flips, while the stable subspace is largely unaffected by the coupling, so that we do not necessarily expect to see an effect of the coupling.

Beyond the simple skew-product structure (\ref{e:skp}), we expect a number of interesting phenomena. In (i)-(iii), we distinguish between uncoupled and unidirectionally linearly  coupled systems. One can easily envision coupling in either direction via nonlinear terms, such as $u_{1,t}=\ldots u_2^\kappa$, $\kappa>1$, and try to derive nonlinear predictions for spreading speeds in the leading edge. Such nonlinear coupling generated slow pushed fronts in the Lotka-Volterra equation; see \cite{lotka}.

\paragraph{Linear Predictions --- Multiple Double Roots and Absolute Spectrum.}

Nonlinear interactions may well couple ``modes'' that are not in strong resonance. Again, some effects become apparent when studying linear systems with general boundary conditions. In \cite{SSabs}, the absolute spectrum was defined through the dispersion relation as follows. For fixed $\lambda\in\C$, order the roots of $d(\lambda,\nu)=0$ by real part, so that $\Re\nu_1\leq\ldots\leq\Re\nu_{2mN}$. The absolute spectrum is defined as
\[
\Sigma_\mathrm{abs}=\{\lambda\in\C\,|\,\Re\nu_{mN}=\Re\nu_{mN+1}\}.
\]
Clearly, pinched double roots belong to the absolute spectrum. For generic boundary conditions, the spectrum on finite but large domains converges to the absolute spectrum setwise, see \cite{SSabs}. 

Comparing to our discussion, the absolute spectrum incorporates possible interactions between roots $(\lambda,\nu)$ and $(\lambda,\nu+\rmi\gamma)$, with equal temporal behavior and equal spatial decay rates, while double roots require strictly equal spatial behavior. Allowing for time-periodic forcing at the boundary, one can also define absolute Floquet spectra \cite{rss,tina}, when 
$(\lambda,\nu)$ and $(\lambda+\rmi\omega,\nu+\rmi\gamma)$. 

We expect that unstable absolute spectra will impact spreading speeds in a similar way as double double roots do. Consider for instance the interaction of a Hopf bifurcation and a pitchfork bifurcation, which would be described by amplitude equations for $A$, the amplitude of the Hopf mode, and $u$, the amplitude of the pitchfork mode. Since frequencies associated with the Hopf mode are nonzero, we will not see double double roots in a coupled system. In an amplitude equation description, one does however average out oscillations with an Ansatz $A\rme^{\rmi\omega t}$, so that in the amplitude equation approximation we would see double double roots, with possible \emph{relevant} nonlinear coupling between Hopf and pitchfork. 

Coming back to the point of view taken in the beginning, Section \ref{s:1.2}, absolute spectra give optimal decay in optimally chosen exponentially weighted spaces. Similarly, relevant and irrelevant double double roots, cases (ii) and (iii) in the previous section, can also be distinguished via exponential weights: in the (irrelevant) case (iii), it is possible to choose exponential weights separately for $u_1$ and $u_2$ so that the linearization is invertible; see also \cite{lotka}. This is not possible in case (ii) since exponential weights require stronger decay in the $u_2$-component, which is incompatible with the direction of coupling. 

We suspect that the possibility of finding exponential weights that stabilize the leading edge gives in most cases a sharp estimate on the actual spreading speed.

\subsection{Conclusion}\label{s:con}

The results in this paper are mostly concerned with the linear theory in the leading edge of invasion processes. Our systematic treatment revealed exotic, ``degenerate'' cases, which however occur in a robust fashion when studying concrete systems, in ecology or in pattern formation. It also revealed a linear rigidity in the formation of patterns, favoring stripes in all linear invasion processes. The discussion in this last section points towards a plethora of interesting nonlinear phenomena. Our discussion of those phenomena is piecemeal at best,  but we expect that the linear theory will be an important ingredient to any systematic theoretical or computational exploration, at the least helping to categorize phenomena.

\end{document}